\documentclass[aps,prd,onecolumn,superscriptaddress,nofootinbib,showpacs]{revtex4}
\usepackage{graphicx}
\usepackage{amssymb,amsmath}
\usepackage{dsfont}

\def\ba{\begin{eqnarray}}
\def\ea{\end{eqnarray}}
\def\be{\begin{equation}}
\def\ee{\end{equation}}

\def\x{\texttt{x}}

\begin{document}

\title{Abelian symmetries in the
  two-Higgs-doublet model with fermions}

\author{P.\ M.\ Ferreira}
\affiliation{Instituto Superior de Engenharia de Lisboa,
    Rua Conselheiro Em\'{\i}dio Navarro,
    1900 Lisboa, Portugal}
\affiliation{Centro de F\'{\i}sica Te\'{o}rica e Computacional,
    Faculdade de Ci\^{e}ncias,
    Universidade de Lisboa,
    Av.\ Prof.\ Gama Pinto 2,
    1649-003 Lisboa, Portugal}
%
%
\author{Jo\~{a}o P.\ Silva}
\affiliation{Instituto Superior de Engenharia de Lisboa,
    Rua Conselheiro Em\'{\i}dio Navarro,
    1900 Lisboa, Portugal}
\affiliation{Centro de F\'{\i}sica Te\'{o}rica de Part\'{\i}culas,
    Instituto Superior T\'{e}cnico,
    P-1049-001 Lisboa, Portugal}

\date{\today}

\begin{abstract}
We classify all possible implementations
of an abelian symmetry in the two-Higgs-doublet model
with fermions.
We identify those symmetries which are consistent non-vanishing
quark masses and a CKM matrix which is not block-diagonal.
Our analysis takes us from a plethora of possibilities down to
246 relevant cases, requiring only 34 distinct matrix forms.
We show that applying $Z_n$ with $n \geq 4$ to the scalar
sector leads to a continuous $U(1)$ symmetry in the whole
Lagrangian.
Finally,
we address the possibilities of spontaneous
CP violation and of natural suppression of the
flavour changing neutral currents. We explain why our
work is relevant even for non-abelian symmetries. 
\end{abstract}


\maketitle

\section{\label{sec:intro}Introduction}

The least known aspect of the electroweak interactions is
its scalar sector.
In the Standard Model (SM) there is only one Higgs
but,
although this is an economical choice,
there is no fundamental reason for nature to adopt it.
Ultimately,
the number of Higgs fields,
like the number of fermion families before it,
must be assessed experimentally.
Partly for this reason,
there has been a great interest in multi-Higgs models.
This is also due to the fact that many interesting new
effects arise,
such as the presence of charged scalars,
the possibility for CP violation in the scalar sector,
and the possibility for spontaneous CP violation,
to name a few.

One problem with multi-Higgs models is that they involve
many more parameters than needed in the SM.
This problem can be tamed by invoking discrete symmetries.
A complete classification of the impact of
discrete and continuous symmetries in the
scalar sector of the two Higgs doublet model (THDM)
has been discussed in the literature \cite{Ivanov_N=2,us},
and some incursions exist into theories with more than
two Higgs doublets \cite{FS1,Ivanov_N=3}.
There are also several articles discussing
specific implementations of discrete symmetries
in both the scalar and fermion sectors,
but no complete classification exists.
This is the problem we tackle here.

This article is organized as follows.
In section~\ref{sec:notation} we
introduce our notation and show the impact
that a choice of abelian symmetries in
the scalar and fermion sectors has on the
Yukawa matrices.
A priori there are $3^{18}$ possibilities.
In section~\ref{sec:classify} we show how
simple experimental considerations,
such as the absence of massless quarks and the
non-block-diagonal nature of the CKM matrix can
be used to curtail this number down to $246$.
Up to permutations,
these involve only 34 forms of Yukawa matrices,
which we show explicitly.
Since any finite discrete group has an abelian
sub-group,
our classification is important even for those
considering non-abelian family symmetries.
We present two important results in section~\ref{sec:important}.
Our classification is then used to address two questions:
whether one can have spontaneous CP violation,
in section~\ref{sec:CPV};
and whether one can relate the flavour changing neutral current
interactions with the CKM matrix,
in section~\ref{sec:FCNSI}.
We draw our conclusions in
section~\ref{sec:conclusions}.

\section{\label{sec:notation}Notation}

\subsection{The Lagrangian}

Let us consider a $SU(2) \otimes U(1)$ gauge theory with
two hypercharge-one Higgs-doublets, denoted by $\Phi_a$,
where $a=1,2$.
The scalar potential may be written as
\begin{eqnarray}
- \mathcal{L}_\mathrm{H} =
&=&
Y_{ab} (\Phi_a^\dagger \Phi_b) +
\tfrac{1}{2}
Z_{ab,cd} (\Phi_a^\dagger \Phi_b) (\Phi_c^\dagger \Phi_d),
\label{VH2}
\end{eqnarray}
where Hermiticity implies
\begin{eqnarray}
Y_{ab} &=& Y_{ba}^\ast,
\nonumber\\
Z_{ab,cd} \equiv Z_{cd,ab} &=& Z_{ba,dc}^\ast.
\label{hermiticity_coefficients}
\end{eqnarray}
Minimization of this potential leads to the vacuum expectation values (vevs)
$\langle \Phi_a \rangle = v_a$.

The theory contains also 3 families of
left-handed quark doublets ($q_L$),
right-handed down-type quarks ($n_R$),
and
right-handed up-type quarks ($p_R$).
For the most part,
we will ignore the leptonic sector,
since the analysis would be similar.
The Yukawa Lagrangian may be written as
\be
\label{yuk}
\mathcal{L}_\mathrm{Y} =
- \bar q_L \left[
\left( \Gamma_1 \Phi_1 + \Gamma_2 \Phi_2 \right) n_R
+
\left( \Delta_1 \tilde \Phi_1 + \Delta_2 \tilde \Phi_2 \right) p_R
\right]
+ \mathrm{H.c.},
\ee
where
$\tilde \Phi_k \equiv i \tau_2 \Phi_k^\ast$,
and $q_L$,
$n_R$,
and $p_R$ are 3-vectors
in flavour space.
The $3 \times 3$ matrices
$\Gamma_k$,
$\Delta_k$,
contain the complex Yukawa couplings
to the right-handed down-type quarks and up-type quarks,
respectively.

\subsection{Basis transformations}

The Lagrangian can be rewritten in terms of new fields
obtained from the original ones by simple basis transformations
\begin{eqnarray}
\Phi_a
& \rightarrow &
\Phi_a^\prime = U_{ab}\ \Phi_b,
\nonumber\\
q_L
& \rightarrow &
q^\prime_L = U_L\ q_L,
\nonumber\\
n_R
& \rightarrow &
n^\prime_R = U_{nR}\ n_R,
\nonumber\\
p_R
& \rightarrow &
p^\prime_R = U_{pR}\ p_R,
\label{basis-transf}
\end{eqnarray}
where $U\in U(2)$ is a $2 \times 2$ unitary matrix,
while $\left\{ U_L, U_{nR}, U_{pR} \right\} \in U(3)$ are $3 \times 3$ unitary matrices.
Under these unitary basis transformations,
the gauge-kinetic terms are unchanged,
but the coefficients $Y_{ab}$ and $Z_{ab,cd}$ are transformed as
\begin{eqnarray}
Y_{ab} & \rightarrow &
Y^\prime_{ab} =
U_{a \alpha}\ Y_{\alpha \beta}\ U_{b \beta}^\ast ,
\label{Y-transf}
\\
Z_{ab,cd} & \rightarrow &
Z^\prime_{ab,cd} =
U_{a\alpha}\, U_{c \gamma}\
Z_{\alpha \beta,\gamma \delta}\ U_{b \beta}^\ast \, U_{d \delta}^\ast,
\label{Z-transf}
\end{eqnarray}
while the Yukawa matrices change as
\begin{eqnarray}
\Gamma_a
& \rightarrow &
\Gamma^\prime_a =
U_L\ \Gamma_\alpha\ U_{nR}^\dagger\ \left( U^\dagger \right)_{ \alpha a}
\nonumber\\
\Delta_a
& \rightarrow &
\Delta^\prime_a =
U_L\ \Delta_\alpha\ U_{pR}^\dagger\ \left( U^\top \right)_{ \alpha a}.
\end{eqnarray}
Notice that we have kept the notation of showing explicitly
the indices in scalar-space,
while using matrix formulation for the quark flavour
spaces.
The basis transformations may be utilized in order to absorb
some of the degrees of freedom of
$Y$, $Z$, $\Gamma$, and/or $\Delta$,
which implies that not all parameters in the Lagrangian
have physical significance.

\subsection{\label{subsec:symmetries}Symmetries in the THDM}

We will now assume that the Lagrangian is invariant under the
symmetry
\begin{eqnarray}
\Phi_a
& \rightarrow &
\Phi_a^S
= S_{ab}\ \Phi_b,
\nonumber\\*[1mm]
q_L
& \rightarrow &
q_L^S
= S_L\ q_L,
\nonumber\\*[1mm]
n_R
& \rightarrow &
n_R^S
= S_{nR}\ n_R,
\nonumber\\*[1mm]
p_R
& \rightarrow &
p_R^S
= S_{pR}\ p_R,
\label{S-transf-symmetry}
\end{eqnarray}
where $S \in U(2)$,
while $\left\{ S_L, S_{nR}, S_{pR} \right\} \in U(3)$.
As a result of this symmetry,
\begin{eqnarray}
Y_{a b} & = &
S_{a \alpha}\ Y_{\alpha \beta}\ S_{b \beta}^\ast ,
\label{Y-S}
\\
Z_{ab,cd} & = &
S_{a \alpha}\, S_{c \gamma}\
Z_{\alpha \beta, \gamma \delta}\ S_{b \beta}^\ast \, S_{d \delta}^\ast ,
\label{Z-S}
\\
\Gamma_a & = &
S_L\ \Gamma_\alpha\ S_{nR}^\dagger\ \left( S^\dagger \right)_{\alpha a} ,
\label{Gamma-S}
\\
\Delta_a & = &
S_L\ \Delta_\alpha\ S_{pR}^\dagger\ \left( S^\top \right)_{\alpha a}.
\label{Delta-S}
\end{eqnarray}

Under the basis transformation of Eq.~(\ref{basis-transf}),
the specific form of the symmetry
in Eq.~\eqref{S-transf-symmetry} is altered as
\begin{eqnarray}
S^\prime &=& U\ S\ U^\dagger ,
\label{S-prime}
\\
S^\prime_L &=& U_L\ S_L\ U^\dagger_L ,
\label{S_L-prime}
\\
S^\prime_{nR} &=& U_{nR}\ S_{nR}\ U^\dagger_{nR} ,
\label{S_nR-prime}
\\
S^\prime_{pR} &=& U_{pR}\ S_{pR}\ U^\dagger_{pR} .
\label{S_pR-prime}
\end{eqnarray}
Suppose that one has chosen to apply the symmetry
$ \left\{S, S_L, S_{nR}, S_{pR} \right\} $ in some basis.
By a judicious choice of
$ \left\{U, U_L, U_{nR}, U_{pR} \right\} $
one may bring the symmetry into the form
\begin{eqnarray}
S
&=&
\textrm{diag} \left\{ e^{i \theta_1}, e^{i \theta_2} \right\},
\label{S-simple1}
\\
S_L
&=&
\textrm{diag} \left\{ e^{i \alpha_1}, e^{i \alpha_2}, e^{i \alpha_3} \right\},
\label{S_L-simple1}
\\
S_{nR}
&=&
\textrm{diag} \left\{ e^{i \beta_1}, e^{i \beta_2}, e^{i \beta_3} \right\},
\label{S_nR-simple1}
\\
S_{pR}
&=&
\textrm{diag} \left\{ e^{i \gamma_1}, e^{i \gamma_2}, e^{i \gamma_3} \right\}.
\label{S_pR-simple1}
\end{eqnarray}

What about global phases?
Clearly,
an overall phase change has no effect on the symmetry.
For example,
taking $ U = e^{i \theta} \mathds{1}_2$,
leaves $S^\prime = S$.
However,
it is easy to see from
Eqs.~(\ref{Y-S})--(\ref{Delta-S})
that the symmetry
\be
\tilde S = e^{i \tilde{\theta}} S,
\ \
\tilde S_L = e^{i \tilde{\alpha}} S_L,
\ \
\tilde S_{nR} = e^{i \tilde{\beta}} S_{nR},
\ \
\tilde S_{pR} = e^{i \tilde{\gamma}} S_{pR},
\ee
imposes the same restrictions on the
Lagrangian as the symmetry
$ \left\{S, S_L, S_{nR}, S_{pR} \right\} $,
as long as
\begin{equation}
e^{i (\tilde{\beta} - \tilde{\alpha} - \tilde{\theta})} = 1
\ \ \ \
\textrm{and}
\ \ \ \
e^{i (\tilde{\gamma} - \tilde{\alpha} + \tilde{\theta})} = 1.
\end{equation}
This can be used to bring Eqs.~(\ref{S-simple1})--(\ref{S_pR-simple1}) into the form
\begin{eqnarray}
S
&=&
\textrm{diag} \left\{ 1, e^{i \theta} \right\},
\label{S-simple2}
\\
S_L
&=&
\textrm{diag} \left\{ e^{i \alpha_1}, e^{i \alpha_2}, e^{i \alpha_3} \right\},
\ \ \textrm{with} \ \ \alpha_1=0,
\label{S_L-simple2}
\\
S_{nR}
&=&
\textrm{diag} \left\{ e^{i \beta_1}, e^{i \beta_2}, e^{i \beta_3} \right\},
\label{S_nR-simple2}
\\
S_{pR}
&=&
\textrm{diag} \left\{ e^{i \gamma_1}, e^{i \gamma_2}, e^{i \gamma_3} \right\}.
\label{S_pR-simple2}
\end{eqnarray}

For $\theta = \pi$,
$S=\textrm{diag}(1,-1)$ leads to the usual $Z_2$ Higgs potential.
Any other value of $0 < \theta < 2 \pi$,
leads to the full $U(1)$ symmetric Higgs potential.
For example,
with $\theta=2 \pi/3$,
$S^3 = \mathds{1}_2$,
and a $Z_3$ symmetry is imposed on the scalar fields.
Nevertheless,
because the scalar potential only has quadratic and quartic
terms,
the resulting Higgs potential has the full $U(1)$
Peccei-Quinn symmetry \cite{FS1}.
If this symmetry is broken spontaneously by the vacuum,
we will have massless particles.
As a result, great care must be taken when imposing
what may look like discrete symmetries in multi-Higgs models.
Substituting Eqs.~(\ref{S-simple1})--(\ref{S_pR-simple1}) in
Eqs.~(\ref{Gamma-S}) and (\ref{Delta-S}),
we find
\begin{eqnarray}
\left( \Gamma_a \right)_{ij}
&=&
e^{i (\alpha_i - \beta_j - \theta_a)} \left( \Gamma_a \right)_{ij},
\label{Gamma-S2}
\\
\left( \Delta_a \right)_{ij}
&=&
e^{i (\alpha_i - \gamma_j + \theta_a)} \left( \Delta_a \right)_{ij},
\label{Delta-S2}
\end{eqnarray}
where \textit{no sum} over $i$ and $j$ is intended on the right-hand sides.
For the simplified form in Eq.~(\ref{S-simple2}) we set $\theta_1=0$ and $\theta_2=\theta$.
Furthermore,
we will always take $\theta \neq 0\ (\textrm{mod } 2 \pi)$,
since we are only interested in symmetries which \textit{do transform the scalar fields}.
It will prove useful to keep $\alpha_1$ explicitly,
bearing in mind that it can be set equal to zero without
loss of generality.
These equations constitute our starting point for what follows.

\subsection{\label{subsec:classify}Preliminary constraints on the Yukawa matrices}

We will concentrate first on the down-type Yukawa matrices $\Gamma_a$.
Given a symmetry written in the form of
Eqs.~(\ref{S-simple2})--(\ref{S_pR-simple2})
we conclude from Eq.~\eqref{Gamma-S2} that
\begin{itemize}
\item $\left( \Gamma_1 \right)_{ij}$ can take any value if $\theta_{ij} = 0$;
\item $\left( \Gamma_1 \right)_{ij} = 0$ if $\theta_{ij} \neq 0$;
\item $\left( \Gamma_2 \right)_{ij}$ can take any value if $\theta_{ij} = \theta$;
\item $\left( \Gamma_2 \right)_{ij} = 0$ if $\theta_{ij} \neq \theta$;
\end{itemize}
where we have defined
\begin{equation}
\theta_{ij} = \alpha_i - \beta_j.
\label{theta_ij}
\end{equation}
We conclude that,
for a matrix $S$ characterized by a given $\theta \neq 0$,
there are only three possibilities:
\begin{enumerate}
\item $\theta_{ij} = 0\ \ \ \ \ \
\Longrightarrow\ \ \  \left( \Gamma_1 \right)_{ij} =$ any \hspace{2.5mm} and \hspace{2.5mm}
$\left( \Gamma_2 \right)_{ij}= 0$ ;
\item $\theta_{ij} = \theta\ \ \ \ \ \, \,
\Longrightarrow\ \ \  \left( \Gamma_1 \right)_{ij}= 0$ \hspace{6mm}  and \hspace{2.5mm}
$\left( \Gamma_2 \right)_{ij} =$ any;
\item $\theta_{ij} \neq 0, \theta\ \ \
\Longrightarrow\ \ \  \left( \Gamma_1 \right)_{ij} = 0 = \left( \Gamma_2 \right)_{ij}$.
\end{enumerate}
All conditions on $\theta_{ij}$ are mod$(2\pi)$.
Noticing that only five $\theta_{ij}$ are independent,
we will take these to be
$\theta_{11}$, $\theta_{12}$, $\theta_{13}$,
$\theta_{21}$, and $\theta_{31}$.
Then,
\begin{eqnarray}
\theta_{22} =  \theta_{21} + \theta_{12} - \theta_{11},
&\hspace{2ex}&
\theta_{23} =  \theta_{21} + \theta_{13} - \theta_{11},
\nonumber\\
\theta_{32} =  \theta_{31} + \theta_{12} - \theta_{11},
&\hspace{2ex}&
\theta_{33} =  \theta_{31} + \theta_{13} - \theta_{11}.
\label{4theta}
\end{eqnarray}
For each $\theta \neq 0$,
we must only consider five $\theta_{ij}$.
The possibilities $\theta_{ij}=0$ and $\theta_{ij}=\theta$
are simple to enumerate.
Unfortunately,
the impact of $\theta_{ij} \neq 0, \theta$ depends
on the exact value of $\theta_{ij}$. 
Thus, there are far more than the $3^5$ possibilities one 
might naively expect. 
For example,
choosing
$\{ \theta_{11}, \theta_{12}, \theta_{21} \}
= \{ 7 \theta, 2 \theta, 2 \theta\}$
and $\theta = \sqrt{2} \pi$,
we conclude that the $(1,1)$, $(1,2)$,
and $(2,1)$ entries of $\Gamma_1$ and $\Gamma_2$
matrices vanish, as do the $(2,2)$ entries.
In contrast,
choosing
$\{ \theta_{11}, \theta_{12}, \theta_{21} \}
= \{ 4 \theta, 2 \theta, 2 \theta\}$ and $\theta = \sqrt{2} \pi$
we conclude that the $(1,1)$, $(1,2)$,
and $(2,1)$ entries of $\Gamma_1$ and $\Gamma_2$
matrices vanish,
but the $(2,2)$ entry of $\Gamma_1$ need not vanish~\footnote{Notice
that the freedom to choose $\alpha_1=0$ does not
reduce the number of possibilities.
}.

Some possibilities are trivially inconsistent with experiment.
For example,
choosing
$\{ \theta_{11}, \theta_{12}, \theta_{13}, \theta_{21}, \theta_{31} \}
= \{ 0, \theta, \theta, \theta, \theta \}$,
then the matrix
\begin{equation}
\Theta = \left\{ \theta_{ij} \right\}
\label{Theta-matrix}
\end{equation}
becomes
\begin{equation}
\Theta =
\left[
\begin{array}{ccc}
0 & \theta & \theta\\
\theta & 2 \theta & 2 \theta\\
\theta & 2 \theta & 2 \theta
\end{array}
\right].
\end{equation}
For $\theta \neq 0$ (identity operation) and $\theta \neq \pi$ (usual $Z_2$ symmetry),
we are lead to Yukawa matrices of the form
\begin{equation}
\Gamma_1 =
\left[
\begin{array}{ccc}
a_{11} & 0 & 0\\
0 & 0 & 0\\
0 & 0 & 0
\end{array}
\right],
\hspace{2ex}
\Gamma_2 =
\left[
\begin{array}{ccc}
0 & b_{12} & b_{13}\\
b_{21} & 0 & 0\\
b_{31} & 0 & 0
\end{array}
\right].
\label{example1}
\end{equation}
Upon spontaneous electroweak symmetry breaking,
the down-type quark mass matrix will arise from the
bi-diagonalization of
\begin{equation}
v_1 \Gamma_1 + v_2 \Gamma_2
=
\left[
\begin{array}{ccc}
v_1 a_{11} & v_2 b_{12} & v_2 b_{13}\\
v_2 b_{21} & 0 & 0\\
v_ 2 b_{31} & 0 & 0
\end{array}
\right],
\end{equation}
whose determinant is zero.
As a result,
this model would lead to one massless quark,
which is ruled out by experiment.
Notice that choosing $\{ \theta_{11}, \theta_{12}, \theta_{13}, \theta_{21}, \theta_{31} \}
= \{ \theta, 0, 0, 0, 0 \}$ would lead to
Yukawa matrices of the form
\begin{equation}
\Gamma_1 =
\left[
\begin{array}{ccc}
0 & b_{12} & b_{13}\\
b_{21} & 0 & 0\\
b_{31} & 0 & 0
\end{array}
\right],
\hspace{2ex}
\Gamma_2 =
\left[
\begin{array}{ccc}
a_{11} & 0 & 0\\
0 & 0 & 0\\
0 & 0 & 0
\end{array}
\right].
\label{example2}
\end{equation}
This is the same as Eq.~\eqref{example1},
with the substitution $\Phi_1 \leftrightarrow \Phi_2$.
Said otherwise,
these possibilities represent the same model.
The interchange $\Phi_1 \leftrightarrow \Phi_2$ cuts down the
number of distinct models by almost a factor of two.

An old model by Lavoura \cite{Lav92} had
\begin{eqnarray}
S
&=&
\textrm{diag} \left\{ 1 , -1 \right\},
\nonumber\\
S_L
&=&
\textrm{diag} \left\{ 1, 1, 1 \right\},
\nonumber\\
S_{nR}
&=&
\textrm{diag} \left\{ 1, 1, -1 \right\},
\nonumber\\
S_{pR}
&=&
\textrm{diag} \left\{ 1, 1, 1 \right\}.
\label{Lav-model}
\end{eqnarray}
Thus
\begin{equation}
\Theta =
\left[
\begin{array}{ccc}
0 & 0 & \theta\\
0 & 0 & \theta\\
0 & 0 & \theta
\end{array}
\right],
\label{Theta-Lav-model}
\end{equation}
leading to Yukawa matrices of the form
\begin{equation}
\Gamma_1 =
\left[
\begin{array}{ccc}
a_{11} & a_{12} & 0\\
a_{21} & a_{22} & 0\\
a_{31} & a_{32} & 0
\end{array}
\right],
\hspace{2ex}
\Gamma_2 =
\left[
\begin{array}{ccc}
0 & 0 & b_{13}\\
0 & 0 & b_{23}\\
0 & 0 & b_{33}
\end{array}
\right].
\label{Yuk-Lav-model}
\end{equation}
A model where
\begin{equation}
\Theta =
\left[
\begin{array}{ccc}
0 & \theta & 0\\
0 & \theta & 0\\
0 & \theta & 0
\end{array}
\right],
\end{equation}
will be indistinguishable from Lavoura's model,
as will a model where the $\theta$s move to the first column.
Such permutations will further cut down the number of
distinct models.

For the up-type sector we define
\begin{equation}
\bar \theta_{ij} = \alpha_i - \gamma_j.
\label{bartheta_ij}
\end{equation}
As before,
for a matrix $S$ characterized by a given $\theta \neq 0$,
there are only three possibilities:
\begin{enumerate}
\item $\bar \theta_{ij} = 0 \hspace{9mm}
\Longrightarrow\ \ \  \left( \Delta_1 \right)_{ij} =$ any \hspace{2.5mm} and \hspace{2.5mm}
$\left( \Delta_2 \right)_{ij}= 0$ ;
\item $\bar \theta_{ij} = -\theta \hspace{6.3mm}
\Longrightarrow\ \ \  \left( \Delta_1 \right)_{ij}= 0$ \hspace{6mm}  and \hspace{2.5mm}
$\left( \Delta_2 \right)_{ij} =$ any;
\item $\bar \theta_{ij} \neq 0, -\theta \hspace{3mm}
\Longrightarrow\ \ \  \left( \Delta_1 \right)_{ij} = 0 = \left( \Delta_2 \right)_{ij}$.
\end{enumerate}
All conditions on $\bar \theta_{ij}$ are mod$(2\pi)$.
Clearly we can choose independently $\bar \theta_{11}$,
$\bar \theta_{12}$,
and $\bar \theta_{13}$,
and then
\begin{eqnarray}
\bar \theta_{21} =  \theta_{21} - \theta_{11} + \bar \theta_{11}
&
\hspace{2ex}
\bar \theta_{22} =  \theta_{21} - \theta_{11} + \bar \theta_{12},
\hspace{2ex}
&
\bar \theta_{23} =  \theta_{21} - \theta_{11} + \bar \theta_{13},
\nonumber\\
\bar \theta_{31} =  \theta_{31} - \theta_{11} + \bar \theta_{11}
&
\hspace{2ex}
\bar \theta_{32} =  \theta_{31} - \theta_{11} + \bar \theta_{12},
\hspace{2ex}
&
\bar \theta_{33} =  \theta_{31} - \theta_{11} + \bar \theta_{13}.
\label{6bartheta}
\end{eqnarray}
There are 9 entries in the down-type Yukawa matrices.
For each there are only three possibilities
(the entry exists in $\Gamma_1$ but not in $\Gamma_2$;
the entry exists in $\Gamma_2$ but not in $\Gamma_1$;
the entry does not exist in either).
The same occurs for the up-type Yukawa matrices.
As a result,
we would have potentially $3^{18}$ possibilities.
But, as we have illustrated above,
interchange and permutations help cut this number down.
More importantly,
many of the models entail massless quarks,
a diagonal CKM matrix,
or other inconsistencies with experiment.
These are ruled out. This is what we turn to next.

\section{\label{sec:classify}Model Classification}

\subsection{\label{subsec:left}The left-space}

The left-handed space (where the left-handed quark doublets live)
is rather constrained because it affects the down-type quark mass matrix,
the up-type quark mass matrix, and also the CKM matrix.
The quark mass matrices are obtained by bi-diagonalizing
the matrices
\begin{eqnarray}
\Gamma
&\equiv &
v_1 \Gamma_1 + v_2 \Gamma_2,
\label{Gamma_d}\\
\Delta
&\equiv &
v_1^\ast \Delta_1 + v_2^\ast \Delta_2,
\label{Delta}.
\end{eqnarray}
whose two indices live in different spaces.
But both indices of the hermitian matrices
\begin{eqnarray}
H_d
&\equiv &
\Gamma \Gamma^\dagger
=
|v_1|^2 \Gamma_1 \Gamma_1^\dagger
+ |v_2|^2 \Gamma_2 \Gamma_2^\dagger
+ v_1 v_2^\ast \Gamma_1 \Gamma_2^\dagger
+ v_1^\ast v_2 \Gamma_2 \Gamma_1^\dagger
\label{H_d}\\
H_u
&\equiv &
\Delta \Delta^\dagger
=
|v_1|^2 \Delta_1 \Delta_1^\dagger
+ |v_2|^2 \Delta_2 \Delta_2^\dagger
+ v_1^\ast v_2 \Delta_1 \Delta_2^\dagger
+ v_1 v_2^\ast \Delta_2 \Delta_1^\dagger
\label{H_u}
\end{eqnarray}
live on the left-space.
These matrices can be diagonalized through unitary matrices
$V_{dL}$ and $V_{uL}$ as
\begin{eqnarray}
V_{dL} H_d V_{dL}^\dagger = D_d^2 = \textrm{diag} \{ m_d^2, m_s^2, m_b^2 \},
\\
V_{uL} H_u V_{uL}^\dagger = D_u^2 = \textrm{diag} \{ m_u^2, m_c^2, m_t^2 \},
\end{eqnarray}
where $V=V_{uL} V_{dL}^\dagger$ is the CKM matrix.

We may now see the impact of the symmetry on the
left-space and how it affects the quark masses and mixings.
We start from Eq.~(\ref{Gamma-S}) in the form
\begin{eqnarray}
\Gamma_1 & = &
S_L\ \Gamma_1\ S_{nR}^\dagger,
\nonumber\\
\Gamma_2 & = &
S_L\ \Gamma_2\ S_{nR}^\dagger\ e^{-i \theta},
\end{eqnarray}
which,
using the simplified form of $S_L$ in Eq.~\eqref{S_L-simple1},
we can combine into
\begin{eqnarray}
\Gamma_1 \Gamma_1^\dagger =
S_L\ \Gamma_1 \Gamma_1^\dagger\ S_L^\dagger
&=&
\left[
\begin{array}{ccc}
A_{11} & A_{12}\ e^{i \alpha_{12}} & A_{13}\ e^{-i \alpha_{31}}\\
A_{21} \ e^{-i \alpha_{12}}& A_{22} & A_{23}\ e^{i \alpha_{23}}\\
A_{31}\ e^{i \alpha_{31}} & A_{32}\ e^{-i \alpha_{23}} & A_{33}
\end{array}
\right],
\\*[3mm]
\Gamma_2 \Gamma_2^\dagger =
S_L\ \Gamma_2 \Gamma_2^\dagger\ S_L^\dagger
&=&
\left[
\begin{array}{ccc}
B_{11} & B_{12}\ e^{i \alpha_{12}} & B_{13}\ e^{-i \alpha_{31}}\\
B_{21} \ e^{-i \alpha_{12}}& B_{22} & B_{23}\ e^{i \alpha_{23}}\\
B_{31}\ e^{i \alpha_{31}} & B_{32}\ e^{-i \alpha_{23}} & B_{33}
\end{array}
\right],
\\*[3mm]
\Gamma_1 \Gamma_2^\dagger =
S_L\ \Gamma_1 \Gamma_2^\dagger\ S_L^\dagger\ e^{i \theta}
&=&
\left[
\begin{array}{ccc}
C_{11}\ e^{i \theta} & C_{12}\ e^{i (\alpha_{12} + \theta)} & C_{13}\ e^{-i (\alpha_{31} - \theta)}\\
C_{21} \ e^{-i (\alpha_{12} - \theta)}& C_{22}\ e^{i \theta} & C_{23}\ e^{i (\alpha_{23} + \theta)}\\
C_{31}\ e^{i (\alpha_{31} + \theta)} & C_{32}\ e^{-i (\alpha_{23} - \theta)} & C_{33}\ e^{i \theta}
\end{array}
\right],
\\*[3mm]
\Gamma_2 \Gamma_1^\dagger =
S_L\ \Gamma_2 \Gamma_1^\dagger\ S_L^\dagger\ e^{-i \theta}
&=&
\left[
\begin{array}{ccc}
D_{11}\ e^{- i \theta} & D_{12}\ e^{i (\alpha_{12} - \theta)} & D_{13}\ e^{-i (\alpha_{31} + \theta)}\\
D_{21} \ e^{-i (\alpha_{12} + \theta)}& D_{22}\ e^{- i \theta} & D_{23}\ e^{i (\alpha_{23} - \theta)}\\
D_{31}\ e^{i (\alpha_{31} - \theta)} & D_{32}\ e^{-i (\alpha_{23} + \theta)} & D_{33}\ e^{- i \theta}
\end{array}
\right].
\end{eqnarray}
In the previous four equations,
$A=\Gamma_1 \Gamma_1^\dagger$,
$B=\Gamma_2 \Gamma_2^\dagger$,
$C=\Gamma_1 \Gamma_2^\dagger$,
and $D=\Gamma_2 \Gamma_1^\dagger$,
respectively.
We have defined
\begin{equation}
\alpha_{12} = \alpha_1 - \alpha_2,
\hspace{3ex}
\alpha_{23} = \alpha_2 - \alpha_3,
\hspace{3ex}
\alpha_{31} = \alpha_3 - \alpha_1,
\label{alphas}
\end{equation}
which satisfy
\begin{equation}
\alpha_{12} + \alpha_{23} + \alpha_{31} = 0.
\label{alpha_equation}
\end{equation}
It is easy to see that the up-type Yukawa matrices satisfy
identical equations, with $\theta \rightarrow - \theta$.

We define the set
\begin{equation}
{\cal J} = \{ x :\, x= 0\, (\textrm{mod}\, 2 \pi)
\ \vee\
x= \theta\, (\textrm{mod}\, 2 \pi)
\ \vee\
x= - \theta\, (\textrm{mod}\, 2 \pi)
\}.
\end{equation}
If $\alpha_{12}, \alpha_{23}, \alpha_{31} \notin {\cal J}$,
then the matrices $\Gamma_1 \Gamma_1^\dagger$,
$\Gamma_2 \Gamma_2^\dagger$,
$\Delta_1 \Delta_1^\dagger$,
and $\Delta_2 \Delta_2^\dagger$ are
diagonal,
while all $12$ and $21$ combinations vanish.
As a result,
$H_d$ and $H_u$ are diagonal and the CKM matrix $V$
is unity.
This is ruled out by experiment.
As a result,
at least one $\alpha_{ij}$ must belong to ${\cal J}$.
Let us imagine that
$\alpha_{12} \in {\cal J}$,
while $\alpha_{23}, \alpha_{31} \notin {\cal J}$.
In that case,
$H_d$ and $H_u$ are block diagonal,
so are the matrices $V_{dL}$ and $V_{uL}$,
and so is the CKM matrix $V$.
This is also ruled out by experiment.
We are left with the cases where
\begin{enumerate}
\item one $\alpha_{ij}$ is not in ${\cal J}$,
while the two others are in ${\cal J}$;
\item all $\alpha_{ij}$ belong to ${\cal J}$.
\end{enumerate}
Next we study these cases in detail.

\subsection{\label{subsec:one}Odd one out}

We look at the case where only one $\alpha_{ij}$ is
not in ${\cal J}$.
Let us take this to be
$\alpha_{12} \notin {\cal J}$,
$\alpha_{23}, \alpha_{31} \in {\cal J}$.
It is easy to see that the only possibilities that
satisfy this requirement are
$\{ \alpha_{12}, \alpha_{23}, \alpha_{31} \} = \{ 2 \theta, -\theta, -\theta\}$
and
$\{ \alpha_{12}, \alpha_{23}, \alpha_{31} \} = \{ - 2 \theta, \theta, \theta\}$.
The second possibility arises from the first through the interchange
$\alpha_1 \leftrightarrow \alpha_2$.
These symmetries act on the left of the Yukawa matrices and, thus,
we go from one to the other by simply interchanging the first two
rows of the corresponding Yukawa matrices.
Similarly,
the relevant cases where
$\alpha_{23} \notin {\cal J}$,
$\alpha_{31}, \alpha_{13} \in {\cal J}$,
and
$\alpha_{31} \notin {\cal J}$,
$\alpha_{12}, \alpha_{23} \in {\cal J}$
are related to the case shown here by mere permutations among the
rows of the respective Yukawa matrices.
As a result,
we show only the case
$\{ \alpha_{12}, \alpha_{23}, \alpha_{31} \} = \{ 2 \theta, -\theta, -\theta\}$.
Using Eqs.~(\ref{theta_ij}),
we obtain $\theta_{21}= \theta_{11} - 2 \theta$,
$\theta_{31}= \theta_{11} - \theta$.
From Eqs.~(\ref{4theta}) we get
\begin{equation}
\Theta =
\left[
\begin{array}{ccc}
\theta_{11} & \theta_{12} & \theta_{13}\\
\theta_{11} - 2 \theta & \theta_{12} - 2 \theta & \theta_{13} - 2 \theta\\
\theta_{11} - \theta & \theta_{12} - \theta & \theta_{13} - \theta
\end{array}
\right].
\end{equation}
The entries of this matrix which equal $0\, (\textrm{mod } 2 \pi)$ lead to
corresponding entries in $\Gamma_1$;
those which equal $\theta\, (\textrm{mod } 2 \pi)$ lead to
corresponding entries in $\Gamma_2$;
all others lead to vanishing entries in $\Gamma_1$, $\Gamma_2$,
and, thus, in $\Gamma$.
Recall that $\Gamma$ cannot have a row of zeros
nor a column of zeros;
otherwise there would be a massless quark.
This is a very powerful constraint.
Let us consider the columns first.
Since there must be at least one entry on each column,
we conclude that
$\theta_{1j} \in \{ 0, \theta, 2\theta, 3\theta \}\ (\textrm{mod } 2 \pi)$.
This would seem to lead to $4^3$ possibilities.
However,
if $\theta_{11}=\theta_{12}= \theta_{13}$,
then there would be a (forbidden) row of zeros.
The reason for this is that we are considering the case where
$ 2 \theta = \alpha_{12} \notin {\cal J}$,
implying that $\theta \neq z_1 \pi$ and
$\theta \neq z_2\, 2 \pi/3$ with
$z_1$ and $z_2$ integers
--
keeping the interval $[0, 2 \pi[$,
$\theta \notin \{ 0, 2 \pi/3, \pi, 4 \pi/3\}$.
This means that $- \theta$, $ \pm 2 \theta$,
and $3 \theta$ can never equal $0\, (\textrm{mod } 2 \pi)$,
nor can they equal $\theta\, (\textrm{mod } 2 \pi$).
Consider for example the possibility that
$\theta_{11}=\theta_{12}= \theta_{13}= 3\theta$.
Then, $\Theta$ would have $3 \theta$ on the first row,
$\theta$ on the second row,
and $2 \theta$ on the last row.
Because $3 \theta$ and $2 \theta$ cannot equal 0 nor $\theta$ (mod $2 \pi$),
this would imply that the first and last rows of
$\Gamma_1$, $\Gamma_2$, and $\Gamma$ vanish,
leading to massless quarks.
Also,
possibilities where two $\theta_{ij}$ are equal to $0$ or
to $3 \theta$ lead to a $2 \times 2$ block of zeros in $\Gamma$
(implying massless quarks) and are, thus, excluded.
There remain only eight independent forms for the
$\Gamma_{i}$ matrices ($\theta \notin \{ 0, 2 \pi/3, \pi, 4 \pi/3\}$):
\begin{itemize}
\item $\theta_{11}=\theta_{12}=\theta$,
$\theta_{13}= 2 \theta$\ \ $(\textrm{mod } 2 \pi)$
\be
\Gamma_1 =
\left[
\begin{array}{ccc}
    &    &    \\
    &    & \x \\
 \x & \x &
\end{array}
\right],
\hspace{3ex}
\Gamma_2 =
\left[
\begin{array}{ccc}
 \x & \x &    \\
    &    &    \\
    &    & \x
\end{array}
\right],
\hspace{3ex} \theta \neq 2 \pi/3, \pi, 4 \pi/3;\footnote{Eqs.~(\ref{odd1})--(\ref{odd8}) are invariant
under the symmetries for all $\theta$,
but they are only the most general forms
consistent with the symmetry for
those symmetries where $\theta \neq 2 \pi/3, \pi, 4 \pi/3$.
See section~\ref{subsec:important2} for details.}
\label{odd1}
\ee
\item $\theta_{11}=\theta_{12}=\theta$,
$\theta_{13}= 3 \theta$\ \ $(\textrm{mod } 2 \pi)$
\be
\Gamma_1 =
\left[
\begin{array}{ccc}
    &    & \hspace{1ex}\\
    &    &    \\
 \x & \x &
\end{array}
\right],
\hspace{3ex}
\Gamma_2 =
\left[
\begin{array}{ccc}
 \x & \x &    \\
    &    & \x \\
    &    &
\end{array}
\right],
\hspace{3ex} \theta \neq 2 \pi/3, \pi, 4 \pi/3;
\label{odd2}
\ee
\item $\theta_{11}=\theta_{12}=2 \theta$,
$\theta_{13}= 0$\ \ $(\textrm{mod } 2 \pi)$
\be
\Gamma_1 =
\left[
\begin{array}{ccc}
    &    & \x \\
 \x & \x &    \\
    &    &
\end{array}
\right],
\hspace{3ex}
\Gamma_2 =
\left[
\begin{array}{ccc}
    &    & \hspace{1ex} \\
    &    &    \\
 \x & \x &
\end{array}
\right],
\hspace{3ex} \theta \neq 2 \pi/3, \pi, 4 \pi/3;
\label{odd3}
\ee
\item $\theta_{11}=\theta_{12}=2 \theta$,
$\theta_{13}= \theta$\ \ $(\textrm{mod } 2 \pi)$
\be
\Gamma_1 =
\left[
\begin{array}{ccc}
    &    &    \\
 \x & \x &    \\
    &    & \x
\end{array}
\right],
\hspace{3ex}
\Gamma_2 =
\left[
\begin{array}{ccc}
    &    & \x \\
    &    &    \\
 \x & \x &
\end{array}
\right],
\hspace{3ex} \theta \neq 2 \pi/3, \pi, 4 \pi/3;
\label{odd4}
\ee
\item $\theta_{11}= 0$, $\theta_{12}=2 \theta$,
$\theta_{13}= \theta$\ \ $(\textrm{mod } 2 \pi)$
\be
\Gamma_1 =
\left[
\begin{array}{ccc}
 \x &    &    \\
    & \x &    \\
    &    & \x
\end{array}
\right],
\hspace{3ex}
\Gamma_2 =
\left[
\begin{array}{ccc}
    &    & \x \\
    &    &    \\
 \hspace{1ex} & \x &
\end{array}
\right],
\hspace{3ex} \theta \neq 2 \pi/3, \pi, 4 \pi/3;
\label{odd5}
\ee
\item $\theta_{11}= 0$, $\theta_{12}=3 \theta$,
$\theta_{13}= \theta$\ \ $(\textrm{mod } 2 \pi)$
\be
\Gamma_1 =
\left[
\begin{array}{ccc}
 \x &    &    \\
    & \hspace{1ex} &    \\
    &    & \x
\end{array}
\right],
\hspace{3ex}
\Gamma_2 =
\left[
\begin{array}{ccc}
    &    & \x \\
    & \x &    \\
 \hspace{1ex} &    &
\end{array}
\right],
\hspace{3ex} \theta \neq 2 \pi/3, \pi, 4 \pi/3;
\label{odd6}
\ee
\item $\theta_{11}= 0$, $\theta_{12}=2 \theta$,
$\theta_{13}= 3 \theta$\ \ $(\textrm{mod } 2 \pi)$
\be
\Gamma_1 =
\left[
\begin{array}{ccc}
 \x &    &    \\
    & \x &    \\
    &    & \hspace{1ex}
\end{array}
\right],
\hspace{3ex}
\Gamma_2 =
\left[
\begin{array}{ccc}
 \hspace{1ex} &    &    \\
    &    & \x \\
    & \x &
\end{array}
\right],
\hspace{3ex} \theta \neq 2 \pi/3, \pi, 4 \pi/3;
\label{odd7}
\ee
\item $\theta_{11}= \theta$, $\theta_{12}=2 \theta$,
$\theta_{13}= 3 \theta$\ \ $(\textrm{mod } 2 \pi)$
\be
\Gamma_1 =
\left[
\begin{array}{ccc}
    &    & \hspace{1ex} \\
    & \x &    \\
 \x &    &
\end{array}
\right],
\hspace{3ex}
\Gamma_2 =
\left[
\begin{array}{ccc}
 \x &    &    \\
    &    & \x \\
    & \x &
\end{array}
\right],
\hspace{3ex} \theta \neq 2 \pi/3, \pi, 4 \pi/3.
\label{odd8}
\ee
\end{itemize}
The $\x$ denote an allowed complex entry;
vacant positions mean that the entry is zero.
All other allowed cases with
$\{ \alpha_{12}, \alpha_{23}, \alpha_{31} \} = \{ 2 \theta, -\theta, -\theta\}$
are related to these by permutations
among the columns.
This corresponds to a mere renaming of the
down-type right-handed fields $\{ n_{R1}, n_{R2},n_{R3} \}$,
having no physical significance.
As explained above,
all permutations of the rows
correspond to physically allowed cases
other than
$\{ \alpha_{12}, \alpha_{23}, \alpha_{31} \} = \{ 2 \theta, -\theta, -\theta\}$.
As a result,
all column and row permutations of the matrices
in Eqs.~(\ref{odd1})--(\ref{odd8}) correspond to
physically allowed models;
permutations on columns have no physical effect;
permutations on rows also have no physical effect but 
must be performed simultaneously
on the down-type matrices $\Gamma$ and on the
up-type matrices $\Delta$.

\subsection{\label{subsec:all}All in ${\cal J}$}

We now turn to the cases where
$\alpha_{12}, \alpha_{23}, \alpha_{31} \in {\cal J}$.
This means that each $\alpha_{ij}$ can only take the values
$0$, $\theta$, or $- \theta$ (mod $2 \pi$).
There would seem to be $3^3$ possibilities.
But Eq.~\eqref{alpha_equation} allows us to exclude a few.
For example,
taking $-\alpha_{12}=\alpha_{23}=\alpha_{31}=\theta\ (\textrm{mod } 2 \pi)$
into Eq.~\eqref{alpha_equation} would mean that
$\theta = 0 \ (\textrm{mod } 2 \pi)$,
a case we are not considering since it
corresponds to unconstrained scalar fields:
$\Phi_1 \rightarrow \Phi_1, \Phi_2 \rightarrow \Phi_2$.
There are some cases which are possible only
for specific values of $\theta$.
Postponing those for subsections~\ref{subsubsec:special_pi} and \ref{subsubsec:special_2pi/3},
we are left with the following cases:
i) $\{ \alpha_{12}, \alpha_{23}, \alpha_{31} \} = \{ 0, 0, 0\}$;
ii) $\{ \alpha_{12}, \alpha_{23}, \alpha_{31} \} = \{ 0, - \theta, \theta\}$
(interchanging rows on the Yukawa matrices for this case
one reaches the cases $\{ \alpha_{12}, \alpha_{23}, \alpha_{31} \} = \{ - \theta, \theta, 0 \}$)
and $\{ \alpha_{12}, \alpha_{23}, \alpha_{31} \} = \{ \theta, 0, - \theta \}$);
and iii) $\{ \alpha_{12}, \alpha_{23}, \alpha_{31} \} = \{ 0, \theta, -\theta\}$
(interchanging rows on the Yukawa matrices for this case
one reaches the cases $\{ \alpha_{12}, \alpha_{23}, \alpha_{31} \} = \{ \theta, -\theta, 0 \}$)
and $\{ \alpha_{12}, \alpha_{23}, \alpha_{31} \} = \{ -\theta, 0, \theta \}$).

\subsubsection{\label{subsubsec:4.1}
$\{ \alpha_{12}, \alpha_{23}, \alpha_{31} \} = \{ 0, 0, 0\}$ and any $\theta$}

In this case, $\alpha_1 = \alpha_2 = \alpha_3$
and $\theta_{11}= \theta_{21} = \theta_{31}$,
leading to
\begin{equation}
\Theta =
\left[
\begin{array}{ccc}
\theta_{11} & \theta_{12} & \theta_{13}\\
\theta_{11} & \theta_{12} & \theta_{13}\\
\theta_{11} & \theta_{12} & \theta_{13}
\end{array}
\right].
\end{equation}
Because a column of zeros in both $\Gamma_1$ and $\Gamma_2$
would lead to massless quarks,
we must have $\theta_{1j} \in \{ 0, \theta\}$.
There are $2^3$ possibilities;
each column must exist in either $\Gamma_1$
or $\Gamma_2$.
Ignoring cases which differ only by permutation of the columns,
we are left with the following structures:
\begin{itemize}
\item All $\theta_{1j}$ equal 0
\be
\Gamma_1 =
\left[
\begin{array}{ccc}
 \x & \x & \x \\
 \x & \x & \x \\
 \x & \x & \x
\end{array}
\right],
\hspace{3ex}
\Gamma_2 =
\left[
\begin{array}{ccc}
  \hspace{1ex}  &  \hspace{1ex}  &  \hspace{1ex}  \\
    &    &    \\
    &    &
\end{array}
\right],
\hspace{3ex} \textrm{any } \theta;
\label{4.1-1}
\ee
\item Two $\theta_{1j}$ equal 0
\be
\Gamma_1 =
\left[
\begin{array}{ccc}
 \x & \x & \hspace{1ex} \\
 \x & \x &    \\
 \x & \x &
\end{array}
\right],
\hspace{3ex}
\Gamma_2 =
\left[
\begin{array}{ccc}
  \hspace{1ex}  &  \hspace{1ex}  &  \x  \\
    &    & \x \\
    &    & \x
\end{array}
\right],
\hspace{3ex} \textrm{any } \theta;
\label{4.1-2}
\ee
\item One $\theta_{1j}$ equals 0
\be
\Gamma_1 =
\left[
\begin{array}{ccc}
 \x & \hspace{1ex} & \hspace{1ex} \\
 \x &    &    \\
 \x &    &
\end{array}
\right],
\hspace{3ex}
\Gamma_2 =
\left[
\begin{array}{ccc}
  \hspace{1ex}  &  \x  &  \x  \\
    & \x & \x \\
    & \x & \x
\end{array}
\right],
\hspace{3ex} \textrm{any } \theta.
\label{4.1-3}
\ee
This is the same as Eq.~\eqref{4.1-2},
with the interchange $\Phi_1 \leftrightarrow \Phi_2$.
%
\item No $\theta_{1j}$ equals 0
\be
\Gamma_1 =
\left[
\begin{array}{ccc}
 \hspace{1ex} & \hspace{1ex} & \hspace{1ex} \\
    &    &    \\
    &    &
\end{array}
\right],
\hspace{3ex}
\Gamma_2 =
\left[
\begin{array}{ccc}
 \x &  \x  &  \x  \\
 \x & \x & \x \\
 \x & \x & \x
\end{array}
\right],
\hspace{3ex} \textrm{any } \theta.
\label{4.1-4}
\ee
This is the same as Eq.~\eqref{4.1-1},
with the interchange $\Phi_1 \leftrightarrow \Phi_2$.
\end{itemize}

\subsubsection{\label{subsubsec:4.3}
$\{ \alpha_{12}, \alpha_{23}, \alpha_{31} \} = \{ 0, - \theta, \theta\}$ and any $\theta$}

Here\footnote{
Recall that the cases $\{ \alpha_{12}, \alpha_{23}, \alpha_{31} \} = \{ - \theta, \theta, 0 \}$)
and $\{ \alpha_{12}, \alpha_{23}, \alpha_{31} \} = \{ \theta, 0, - \theta \}$
are obtained from this through permutations on the rows of
the Yukawa matrices.}
$\theta_{21}=\theta_{11}$, $\theta_{31}=\theta_{11} + \theta$,
and
\begin{equation}
\Theta =
\left[
\begin{array}{ccc}
\theta_{11} & \theta_{12} & \theta_{13}\\
\theta_{11} & \theta_{12} & \theta_{13}\\
\theta_{11} + \theta & \theta_{12} + \theta & \theta_{13} + \theta
\end{array}
\right],
\end{equation}
implying that $\theta_{1j} \in \{ 0, \theta, - \theta\}$.
Ignoring cases which differ only by permutation of the columns,
we are left with the following structures:
\begin{itemize}
%
\item $\{ \theta_{11}, \theta_{12}, \theta_{13} \} = \{ 0, 0, 0 \}$
\be
\Gamma_1 =
\left[
\begin{array}{ccc}
 \x & \x & \x \\
 \x & \x & \x \\
    &    &
\end{array}
\right],
\hspace{3ex}
\Gamma_2 =
\left[
\begin{array}{ccc}
    &    &    \\
    &    &    \\
 \x & \x & \x
\end{array}
\right],
\hspace{3ex} \textrm{any } \theta;
\label{4.3-g}
\ee
%
\item $\{ \theta_{11}, \theta_{12}, \theta_{13} \} = \{ 0, 0, \theta \}$
\be
\Gamma_1 =
\left[
\begin{array}{ccc}
 \x & \x & \hspace{1ex} \\
 \x & \x &    \\
    &    &
\end{array}
\right],
\hspace{3ex}
\Gamma_2 =
\left[
\begin{array}{ccc}
    &    & \x \\
    &    & \x \\
 \x & \x &
\end{array}
\right],
\hspace{3ex} \theta \neq \pi;
\label{4.3-h}
\ee
\be
\Gamma_1 =
\left[
\begin{array}{ccc}
 \x & \x &    \\
 \x & \x &    \\
    &    & \x
\end{array}
\right],
\hspace{3ex}
\Gamma_2 =
\left[
\begin{array}{ccc}
    &    & \x \\
    &    & \x \\
 \x & \x &
\end{array}
\right],
\hspace{3ex} \theta = \pi.
\label{4.3-d}
\ee
The cases with $\{ \theta_{11}, \theta_{12}, \theta_{13} \}$ equal to
$\{ \theta, 0, 0 \}$
and
$\{ 0, \theta, 0 \}$
are obtained from these
through column permutations.
%
\item $\{ \theta_{11}, \theta_{12}, \theta_{13} \} = \{ 0, \theta, \theta \}$
\be
\Gamma_1 =
\left[
\begin{array}{ccc}
 \x & \hspace{1ex} & \hspace{1ex} \\
 \x &    &    \\
    &    &
\end{array}
\right],
\hspace{3ex}
\Gamma_2 =
\left[
\begin{array}{ccc}
    & \x & \x \\
    & \x & \x \\
 \x &    &
\end{array}
\right],
\hspace{3ex} \theta \neq \pi;
\label{4.3-i}
\ee
\be
\Gamma_1 =
\left[
\begin{array}{ccc}
 \x &    &    \\
 \x &    &    \\
    & \x & \x
\end{array}
\right],
\hspace{3ex}
\Gamma_2 =
\left[
\begin{array}{ccc}
    & \x & \x \\
    & \x & \x \\
 \x &    &
\end{array}
\right],
\hspace{3ex} \theta = \pi.
\label{4.3-b}
\ee
The cases with $\{ \theta_{11}, \theta_{12}, \theta_{13} \}$ equal to
$\{ \theta, \theta, 0 \}$
and
$\{ \theta, 0, \theta \}$
are obtained from these
through column permutations.
%
\item $\{ \theta_{11}, \theta_{12}, \theta_{13} \} = \{ 0, 0, - \theta \}$
\be
\Gamma_1 =
\left[
\begin{array}{ccc}
 \x & \x &    \\
 \x & \x &    \\
    &    & \x
\end{array}
\right],
\hspace{3ex}
\Gamma_2 =
\left[
\begin{array}{ccc}
    &    & \hspace{1ex} \\
    &    &    \\
 \x & \x &
\end{array}
\right],
\hspace{3ex} \theta \neq \pi.
\label{4.3-c}
\ee
Setting $\theta=\pi$ we re-obtain Eq.~\eqref{4.3-d}.
The cases with $\{ \theta_{11}, \theta_{12}, \theta_{13} \}$ equal to
$\{ 0, -\theta, 0 \}$
and
$\{ -\theta, 0, 0 \}$
are obtained from these
through column permutations.
%
%
\item $\{ \theta_{11}, \theta_{12}, \theta_{13} \} = \{ 0, \theta, - \theta \}$
\be
\Gamma_1 =
\left[
\begin{array}{ccc}
 \x & \hspace{1ex} &    \\
 \x &    &    \\
    &    & \x
\end{array}
\right],
\hspace{3ex}
\Gamma_2 =
\left[
\begin{array}{ccc}
    & \x & \hspace{1ex} \\
    & \x &    \\
 \x &    &
\end{array}
\right],
\hspace{3ex} \theta \neq \pi.
\label{4.3-f}
\ee
Setting $\theta=\pi$ we re-obtain Eq.~\eqref{4.3-b}.
The cases with $\{ \theta_{11}, \theta_{12}, \theta_{13} \}$ equal to
$\{ 0, -\theta, \theta \}$,
$\{ \theta, -\theta, 0 \}$,
$\{ \theta, 0, -\theta \}$,
$\{ -\theta, 0, \theta \}$,
and
$\{ -\theta, \theta, 0 \}$
are obtained from these
through column permutations.
%
%
\item $\{ \theta_{11}, \theta_{12}, \theta_{13} \} = \{ \theta, \theta, - \theta \}$
\be
\Gamma_1 =
\left[
\begin{array}{ccc}
 \hspace{1ex} & \hspace{1ex} &    \\
    &    &    \\
    &    & \x
\end{array}
\right],
\hspace{3ex}
\Gamma_2 =
\left[
\begin{array}{ccc}
 \x & \x & \hspace{1ex} \\
 \x & \x &    \\
    &    &
\end{array}
\right],
\hspace{3ex} \theta \neq \pi;
\label{4.3-e}
\ee
\be
\Gamma_1 =
\left[
\begin{array}{ccc}
    &    &    \\
    &    &    \\
 \x & \x & \x
\end{array}
\right],
\hspace{3ex}
\Gamma_2 =
\left[
\begin{array}{ccc}
 \x & \x & \x \\
 \x & \x & \x \\
    &    &
\end{array}
\right],
\hspace{3ex} \theta = \pi.
\label{4.3-a}
\ee
The cases with $\{ \theta_{11}, \theta_{12}, \theta_{13} \}$ equal to
$\{ \theta, -\theta, \theta \}$,
and
$\{ -\theta, \theta, \theta \}$
are obtained from these
through column permutations.
\end{itemize}
For those wishing to check that all possibilities have been considered,
we refer to the
footnote\footnote{
We have also checked that
\begin{itemize}
%
\item The cases where $\{ \theta_{11}, \theta_{12}, \theta_{13} \}$ equal
$\{ 0, -\theta, -\theta \}$,
$\{ -\theta, 0, -\theta \}$,
and
$\{ -\theta, -\theta, 0 \}$
lead to vanishing quark masses,
if $\theta \neq \pi$,
and to Eq.~\eqref{4.3-b},
if $\theta = \pi$;
%
\item The cases where $\{ \theta_{11}, \theta_{12}, \theta_{13} \}$ equal
$\{ \theta, \theta, \theta \}$
lead to vanishing quark masses,
if $\theta \neq \pi$,
and to Eq.~\eqref{4.3-a},
if $\theta = \pi$;
%
\item The cases where $\{ \theta_{11}, \theta_{12}, \theta_{13} \}$ equal
$\{ \theta, -\theta, -\theta \}$,
$\{ -\theta, \theta, -\theta \}$,
and
$\{ -\theta, -\theta, \theta \}$
lead to vanishing quark masses,
if $\theta \neq \pi$,
and to Eq.~\eqref{4.3-a},
if $\theta = \pi$;
%
\item The cases where $\{ \theta_{11}, \theta_{12}, \theta_{13} \}$ equal
$\{ -\theta, -\theta, -\theta \}$
lead to vanishing quark masses,
if $\theta \neq \pi$,
and to Eq.~\eqref{4.3-a},
if $\theta = \pi$.
\end{itemize}
}.

\subsubsection{\label{subsubsec:4.2}
$\{ \alpha_{12}, \alpha_{23}, \alpha_{31} \} = \{ 0, \theta, -\theta\}$ and any $\theta$}

Here\footnote{
Recall that the cases $\{ \alpha_{12}, \alpha_{23}, \alpha_{31} \} = \{ \theta, -\theta, 0 \}$)
and $\{ \alpha_{12}, \alpha_{23}, \alpha_{31} \} = \{ -\theta, 0, \theta \}$
are obtained from this through permutations on the rows of
the Yukawa matrices.}
$\theta_{21}=\theta_{11}$, $\theta_{31}=\theta_{11} - \theta$,
and
\begin{equation}
\Theta =
\left[
\begin{array}{ccc}
\theta_{11} & \theta_{12} & \theta_{13}\\
\theta_{11} & \theta_{12} & \theta_{13}\\
\theta_{11} - \theta & \theta_{12} - \theta & \theta_{13}- \theta
\end{array}
\right],
\end{equation}
implying that $\theta_{1j} \in \{ 0, \theta, 2 \theta\}$.

Ignoring cases which differ only by permutation of the columns,
we are left with the following structures:
\begin{itemize}
%
\item $\{ \theta_{11}, \theta_{12}, \theta_{13} \} = \{ 0, 0, \theta \}$
\be
\Gamma_1 =
\left[
\begin{array}{ccc}
 \x & \x &    \\
 \x & \x &    \\
    &    & \x
\end{array}
\right],
\hspace{3ex}
\Gamma_2 =
\left[
\begin{array}{ccc}
 \hspace{1ex} & \hspace{1ex} & \x \\
    &    & \x \\
    &    &
\end{array}
\right],
\hspace{3ex} \theta \neq \pi;
\label{4.2.2-notpi}
\ee
Performing $\Phi_1 \leftrightarrow \Phi_2$ and exchanging the
first and third columns on Eq.~\eqref{4.2.2-notpi} we obtain Eq.~\eqref{4.3-i}.
Setting $\theta=\pi$ in this case would lead directly to Eq.~\eqref{4.3-d}.
The cases with $\{ \theta_{11}, \theta_{12}, \theta_{13} \}$ equal to
$\{ 0, \theta, 0 \}$
and
$\{ \theta, 0, 0 \}$
are obtained from these
through column permutations.
%
\item $\{ \theta_{11}, \theta_{12}, \theta_{13} \} = \{ 0, \theta, \theta \}$
\be
\Gamma_1 =
\left[
\begin{array}{ccc}
 \x &    &    \\
 \x &    &    \\
    & \x & \x
\end{array}
\right],
\hspace{3ex}
\Gamma_2 =
\left[
\begin{array}{ccc}
 \hspace{1ex} & \x & \x \\
    & \x & \x \\
    &    &
\end{array}
\right],
\hspace{3ex} \theta \neq \pi;
\label{4.2.3-notpi}
\ee
Performing $\Phi_1 \leftrightarrow \Phi_2$ and exchanging the
first and third columns on Eq.~\eqref{4.2.3-notpi} we obtain Eq.~\eqref{4.3-h}.
Setting $\theta=\pi$ in this case would lead directly to Eq.~\eqref{4.3-b}.
The cases with $\{ \theta_{11}, \theta_{12}, \theta_{13} \}$ equal to
$\{ \theta, 0, \theta \}$
and
$\{ \theta, \theta, 0 \}$
are obtained from these
through column permutations.
%
\item $\{ \theta_{11}, \theta_{12}, \theta_{13} \} = \{ 0, 0, 2 \theta \}$
\be
\Gamma_1 =
\left[
\begin{array}{ccc}
 \x & \x & \hspace{1ex} \\
 \x & \x &    \\
    &    &
\end{array}
\right],
\hspace{3ex}
\Gamma_2 =
\left[
\begin{array}{ccc}
 \hspace{1ex} & \hspace{1ex} &    \\
    &    &    \\
    &    & \x
\end{array}
\right],
\hspace{3ex} \theta \neq \pi;
\label{4.2.4-notpi}
\ee
Performing $\Phi_1 \leftrightarrow \Phi_2$ on Eq.~\eqref{4.2.4-notpi}
we obtain Eq.~\eqref{4.3-e}.
Setting $\theta=\pi$ in this case would lead directly to the
special case of $\theta=\pi$ in Eq.~\eqref{4.3-g}.
The cases with $\{ \theta_{11}, \theta_{12}, \theta_{13} \}$ equal to
$\{ 0, 2 \theta, 0 \}$
and
$\{ 2 \theta, 0, 0 \}$
are obtained from these
through column permutations.
%
\item $\{ \theta_{11}, \theta_{12}, \theta_{13} \} = \{ 0, \theta, 2 \theta \}$
\be
\Gamma_1 =
\left[
\begin{array}{ccc}
 \x &    & \hspace{1ex} \\
 \x &    &    \\
    & \x &
\end{array}
\right],
\hspace{3ex}
\Gamma_2 =
\left[
\begin{array}{ccc}
 \hspace{1ex} & \x &  \\
    & \x &    \\
    &    & \x
\end{array}
\right],
\hspace{3ex} \theta \neq \pi.
\label{4.2.6-notpi}
\ee
Performing $\Phi_1 \leftrightarrow \Phi_2$ and exchanging the
first and second columns on Eq.~\eqref{4.2.6-notpi} we obtain Eq.~\eqref{4.3-f}.
Setting $\theta=\pi$ in this case would lead to Eq.~\eqref{4.3-d},
after interchanging the second and third columns.
The cases with $\{ \theta_{11}, \theta_{12}, \theta_{13} \}$ equal to
$\{ 0, 2 \theta, \theta \}$,
$\{ \theta, 2\theta, 0 \}$,
$\{ \theta, 0, 2\theta \}$,
$\{ 2 \theta, 0, \theta \}$,
and
$\{ 2 \theta, \theta, 0 \}$
are obtained from these
through column permutations.
%
\item $\{ \theta_{11}, \theta_{12}, \theta_{13} \} = \{ \theta, \theta, \theta \}$
\be
\Gamma_1 =
\left[
\begin{array}{ccc}
    &    &    \\
    &    &    \\
 \x & \x & \x
\end{array}
\right],
\hspace{3ex}
\Gamma_2 =
\left[
\begin{array}{ccc}
 \x & \x & \x \\
 \x & \x & \x \\
    &    &
\end{array}
\right],
\hspace{3ex} \textrm{any } \theta.
\label{4.2.7}
\ee
Performing $\Phi_1 \leftrightarrow \Phi_2$ on Eq.~\eqref{4.2.7}
we obtain Eq.~\eqref{4.3-g}.
Notice that the special case of $\theta=\pi$ had
already shown up in Eq.~\eqref{4.3-a}.
%
\item $\{ \theta_{11}, \theta_{12}, \theta_{13} \} = \{ \theta, \theta, 2 \theta \}$
\be
\Gamma_1 =
\left[
\begin{array}{ccc}
    &    &  \hspace{1ex} \\
    &    &    \\
 \x & \x &
\end{array}
\right],
\hspace{3ex}
\Gamma_2 =
\left[
\begin{array}{ccc}
 \x & \x &    \\
 \x & \x &    \\
    &    & \x
\end{array}
\right],
\hspace{3ex} \theta \neq \pi;
\label{4.2.8-notpi}
\ee
Performing $\Phi_1 \leftrightarrow \Phi_2$ on Eq.~\eqref{4.2.8-notpi}
we obtain Eq.~\eqref{4.3-c}.
Setting $\theta=\pi$ in this case would lead to Eq.~\eqref{4.3-b},
after interchanging the first and third columns.
The cases with $\{ \theta_{11}, \theta_{12}, \theta_{13} \}$ equal to
$\{ \theta, 2 \theta, \theta \}$
and
$\{ 2 \theta, \theta, \theta \}$
are obtained from these
through column permutations.
\end{itemize}
For those wishing to check that all possibilities have been considered,
we refer to the
footnote\footnote{
We have also checked that
\begin{itemize}
%
\item The cases where $\{ \theta_{11}, \theta_{12}, \theta_{13} \}$ equal
$\{ 0, 0 , 0 \}$
lead to vanishing quark masses,
if $\theta \neq \pi$,
and to Eq.~\eqref{4.3-g},
if $\theta = \pi$;
%
\item The cases where $\{ \theta_{11}, \theta_{12}, \theta_{13} \}$ equal
$\{ 0, 2\theta, 2\theta \}$,
$\{ 2\theta, 0, 2\theta \}$,
and
$\{ 2\theta, 2\theta, 0 \}$
lead to vanishing quark masses,
if $\theta \neq \pi$,
and to Eq.~\eqref{4.3-g},
if $\theta = \pi$;
%
\item The cases where $\{ \theta_{11}, \theta_{12}, \theta_{13} \}$ equal
$\{ \theta, 2\theta, 2\theta \}$,
$\{ 2\theta, \theta, 2\theta \}$,
and
$\{ 2\theta, 2\theta, \theta \}$
lead to vanishing quark masses,
if $\theta \neq \pi$,
and to Eq.~\eqref{4.3-d},
if $\theta = \pi$;
%
\item The cases where $\{ \theta_{11}, \theta_{12}, \theta_{13} \}$ equal
$\{ 2\theta, 2\theta, 2\theta \}$
lead to vanishing quark masses,
if $\theta \neq \pi$,
and to Eq.~\eqref{4.3-g},
if $\theta = \pi$.
\end{itemize}
}.

\subsubsection{\label{subsubsec:special_pi}Special cases with $\theta = \pi$}

We continue to explore the cases where
each $\alpha_{ij}$ can only take the values
$0$, $\theta$, or $- \theta$ (mod $2 \pi$).
Certain cases are only valid for $\theta = \pi$.
For example,
consider $\alpha_{12} = 0\, (\textrm{mod } 2 \pi)$
and $\alpha_{23} = \alpha_{31} = \theta\, (\textrm{mod } 2 \pi)$.
Taking $\theta \in [0, 2 \pi[$,
this can only happen for $\theta = \pi$,
due to Eq.~\eqref{alpha_equation}.
This forces us to consider the case
$\{ \alpha_{12}, \alpha_{23}, \alpha_{31} \} = \{ 0, \pi, \pi \}$.
The cases
$\{ \alpha_{12}, \alpha_{23}, \alpha_{31} \} = \{ \pi, 0, \pi \}$
and
$\{ \alpha_{12}, \alpha_{23}, \alpha_{31} \} = \{ \pi, \pi, 0 \}$
are obtained from this by permuting the rows on the respective
Yukawa matrices.
In this case,
$\theta_{21}=\theta_{11}$,
$\theta_{31}=\theta_{11}+\pi$,
and
\begin{equation}
\Theta =
\left[
\begin{array}{ccc}
\theta_{11} & \theta_{12} & \theta_{13}\\
\theta_{11} & \theta_{12} & \theta_{13}\\
\theta_{11} + \pi & \theta_{12} + \pi & \theta_{13} + \pi
\end{array}
\right],
\end{equation}
implying that $\theta_{1j} \in \{ 0, \pi\}$.
There are $2^3$ such cases,
all of which lead to a matrix $\Gamma$ where
all entries may be non-vanishing\footnote{Of course,
some entry may be zero by accident.
The point is that this value is not required by
a symmetry of this type and, as such,
it is not invariant under the renormalization group equations.
}.
We continue to ignore cases which differ
only by permutation of the columns.
It is easy to see that we have already considered all possible structures.
Indeed,
when all $\theta_{1j}$ equal $\pi$,
we recover Eq.~(\ref{4.3-g});
when two $\theta_{1j}$ equal $\pi$,
we recover Eq.~(\ref{4.3-d});
when only one $\theta_{1j}$ equals $\pi$,
we recover Eq.~(\ref{4.3-b});
and when no $\theta_{1j}$ equals $\pi$,
we recover Eq.~(\ref{4.3-a}).

\subsubsection{\label{subsubsec:special_2pi/3}Special cases with $\theta = 2 \pi/3$}

We now turn to the last two cases where
each $\alpha_{ij}$ can only take the values
$0$, $\theta$, or $- \theta$ (mod $2 \pi$).
Due to Eq.~\eqref{alpha_equation},
we can have $\alpha_{12}=\alpha_{23}=\alpha_{31}=\pm \theta$
if and only if $\theta = 2 \pi/3$.
The case $\alpha_{12}=\alpha_{23}=\alpha_{31}=-2 \pi/3$
(or, which is the same, $4 \pi/3$)
is obtained by exchanging any two rows of the
Yukawa matrices.
We choose the case
$\alpha_{12}=\alpha_{23}=\alpha_{31}=2 \pi/3$,
implying that
$\theta_{21}=\theta_{11}- 2\pi/3$,
$\theta_{31}=\theta_{11}+ 2\pi/3$,
and
\begin{equation}
\Theta =
\left[
\begin{array}{ccc}
\theta_{11} & \theta_{12} & \theta_{13}\\
\theta_{11} -2\pi/3 & \theta_{12} -2\pi/3 & \theta_{13} - 2\pi/3 \\
\theta_{11} + 2\pi/3 & \theta_{12} + 2\pi/3 & \theta_{13}+ 2\pi/3
\end{array}
\right],
\end{equation}
implying that $\theta_{1j} \in \{ 0, 2\pi/3, -2\pi/3\}$.
Recall that $\theta_{11}=\theta_{12}=\theta_{13}$ is excluded
because it would lead to massless quarks.

Ignoring cases which differ only by permutation of the columns,
we are left with the following structures:
\begin{itemize}
%
\item $\{ \theta_{11}, \theta_{12}, \theta_{13} \} = \{ 0, 0, 2 \pi/3 \}$
\be
\Gamma_1 =
\left[
\begin{array}{ccc}
 \x & \x &    \\
    &    & \x \\
    &    &
\end{array}
\right],
\hspace{3ex}
\Gamma_2 =
\left[
\begin{array}{ccc}
    &    & \x \\
    &    &    \\
 \x & \x &
\end{array}
\right],
\hspace{3ex} \theta = 2\pi/3;
\label{2p3.1}
\ee
%
\item $\{ \theta_{11}, \theta_{12}, \theta_{13} \} = \{ 0, 0, -2 \pi/3 \}$
\be
\Gamma_1 =
\left[
\begin{array}{ccc}
 \x & \x &    \\
    &    &    \\
    &    & \x
\end{array}
\right],
\hspace{3ex}
\Gamma_2 =
\left[
\begin{array}{ccc}
    &    &    \\
    &    & \x \\
 \x & \x &
\end{array}
\right],
\hspace{3ex} \theta = 2\pi/3;
\label{2p3.2}
\ee
%
\item $\{ \theta_{11}, \theta_{12}, \theta_{13} \} = \{ 0, 2\pi/3, 2 \pi/3 \}$
\be
\Gamma_1 =
\left[
\begin{array}{ccc}
 \x &    &    \\
    & \x & \x \\
    &    &
\end{array}
\right],
\hspace{3ex}
\Gamma_2 =
\left[
\begin{array}{ccc}
    & \x & \x \\
    &    &    \\
 \x &    &
\end{array}
\right],
\hspace{3ex} \theta = 2\pi/3;
\label{2p3.3}
\ee
%
\item $\{ \theta_{11}, \theta_{12}, \theta_{13} \} = \{ 0, 2\pi/3, -2 \pi/3 \}$
\be
\Gamma_1 =
\left[
\begin{array}{ccc}
 \x &    &    \\
    & \x &    \\
    &    & \x
\end{array}
\right],
\hspace{3ex}
\Gamma_2 =
\left[
\begin{array}{ccc}
    & \x &    \\
    &    & \x \\
 \x &    &
\end{array}
\right],
\hspace{3ex} \theta = 2\pi/3;
\label{2p3.4}
\ee
%
\item $\{ \theta_{11}, \theta_{12}, \theta_{13} \} = \{ 0, -2\pi/3, -2 \pi/3 \}$
\be
\Gamma_1 =
\left[
\begin{array}{ccc}
 \x &    &    \\
    &    &    \\
    & \x & \x
\end{array}
\right],
\hspace{3ex}
\Gamma_2 =
\left[
\begin{array}{ccc}
    &    &    \\
    & \x & \x \\
 \x &    &
\end{array}
\right],
\hspace{3ex} \theta = 2\pi/3;
\label{2p3.5}
\ee
%
\item $\{ \theta_{11}, \theta_{12}, \theta_{13} \} = \{ 2\pi/3, 2\pi/3, -2 \pi/3 \}$
\be
\Gamma_1 =
\left[
\begin{array}{ccc}
    &    &    \\
 \x & \x &    \\
    &    & \x
\end{array}
\right],
\hspace{3ex}
\Gamma_2 =
\left[
\begin{array}{ccc}
 \x & \x &    \\
    &    & \x \\
    &    &
\end{array}
\right],
\hspace{3ex} \theta = 2\pi/3;
\label{2p3.6}
\ee
%
\item $\{ \theta_{11}, \theta_{12}, \theta_{13} \} = \{ 2\pi/3, -2\pi/3, -2 \pi/3 \}$
\be
\Gamma_1 =
\left[
\begin{array}{ccc}
    &    &    \\
 \x &    &    \\
    & \x & \x
\end{array}
\right],
\hspace{3ex}
\Gamma_2 =
\left[
\begin{array}{ccc}
 \x &    &    \\
    & \x & \x \\
    &    &
\end{array}
\right],
\hspace{3ex} \theta = 2\pi/3;
\label{2p3.7}
\ee
\end{itemize}
Care must be exercised when comparing these
matrices with those shown previously.
Consider, for example,
Eq.~\eqref{2p3.1}.
$\{ \theta_{11}, \theta_{12}, \theta_{13} \} = \{ 0, 0, 2 \pi/3 \}$,
with
$\{ \theta_{21}, \theta_{31} \} =
\{ \theta_{11}-2\pi/3, \theta_{11} + 2\pi/3\}
= \{ -2\pi/3, 2\pi/3\}$.
When might worry about
Eq.~\eqref{4.3-h},
where one can also choose
$\{ \theta_{11}, \theta_{12}, \theta_{13} \} = \{ 0, 0, 2 \pi/3 \}$.
However,
there,
$\{ \theta_{21}, \theta_{31} \} =
\{ \theta_{11}, \theta_{11} + 2\pi/3\}
= \{ 0, 2\pi/3\}$.

\subsection{\label{subsec:up}Yukawa matrices for up-type quarks}

So far, we have only shown the Yukawa matrices for the down-type quarks.
We will now show that it is trivial to get the
Yukawa matrices for the up-type quarks from those
for the down-type quarks.
Let us start from some specific transformation of the left-handed fields,
characterized by $\alpha_{12}$ and $\alpha_{31}$.
From Eqs.~\eqref{4theta} and \eqref{6bartheta} we get
$\theta_{21}=\theta_{11}-\alpha_{12}$,
$\theta_{31}=\theta_{11}+\alpha_{31}$,
so that
\begin{eqnarray}
\Theta &=&
\left[
\begin{array}{ccc}
\theta_{11} & \theta_{12} & \theta_{13}\\
\theta_{11} -\alpha_{12} & \theta_{12} -\alpha_{12}& \theta_{13} -\alpha_{12}\\
\theta_{11} +\alpha_{31}& \theta_{12} +\alpha_{31}& \theta_{13} +\alpha_{31}
\end{array}
\right],
\\
\bar \Theta &=&
\left[
\begin{array}{ccc}
\bar \theta_{11} & \bar \theta_{12} & \bar \theta_{13}\\
\bar \theta_{11} -\alpha_{12} & \bar \theta_{12} -\alpha_{12}& \bar \theta_{13} -\alpha_{12}\\
\bar \theta_{11} +\alpha_{31}& \bar \theta_{12} +\alpha_{31}& \bar \theta_{13} +\alpha_{31}
\end{array}
\right].
\end{eqnarray}
Each entry on the column $j$ of $\Theta$ is of the form $\theta_{1j}+ b$.
We then followed the procedure
\begin{eqnarray}
\theta_{1j} + b = 0\, (\textrm{mod } 2 \pi )
&\ \ \  \Rightarrow\ \ \  &
\textrm{entry is in } \Gamma_1,
\nonumber\\
\theta_{1j} + b = \theta\, (\textrm{mod } 2 \pi )
& \ \ \ \Rightarrow\ \ \  &
\textrm{entry is in } \Gamma_2.
\end{eqnarray}
Let us call $\bar \theta_{1j} = \theta_{1j} - \theta$.
Then, if $\theta_{1j} + b = 0\ (\theta)$,
we find $\bar \theta_{1j} + b = - \theta\ (0)$,
meaning that this is an entry in $\Delta_2$ ($\Delta_1$).
Thus
\begin{eqnarray}
\bar \theta_{1j} + b = (\theta_{1j} - \theta) + b = -\theta\, (\textrm{mod } 2 \pi )
&\ \ \  \Rightarrow\ \ \  &
\textrm{entry is in } \Delta_2,
\nonumber\\
\bar \theta_{1j} + b = (\theta_{1j} - \theta)  + b = 0\, (\textrm{mod } 2 \pi ) \ \
& \ \ \ \Rightarrow\ \ \  &
\textrm{entry is in } \Delta_1.
\end{eqnarray}
The argument goes both ways,
so we can find all cases for the up-type Yukawa matrices $\Delta$
by starting from all cases for the down-type Yukawa matrices $\Gamma$
and performing the following procedure:
\begin{itemize}
\item $\theta_{1j}\ \ \  \longrightarrow\ \ \  \bar \theta_{1j} = \theta_{1j} - \theta$;
\item $\Gamma_1 \ \ \  \longrightarrow\ \ \  \Delta_2$;
\item $\Gamma_2 \ \ \  \longrightarrow\ \ \  \Delta_1$.
\end{itemize}
Of course,
one can shuffle differently the columns of
$\{ \Gamma_1, \Gamma_2 \}$ and $\{ \Delta_2, \Delta_1 \}$,
since they live on different right-handed spaces.

\subsection{\label{subsec:counting}Counting the number of models}
\label{sec:2pi3}

The only purpose of our parameter counting is to show the enormous amount
of cases which have been killed by the simple requirements
that there be no massless quarks and that the CKM matrix not be block diagonal.
As pointed out at the end of section~\ref{sec:notation},
there are potentially $3^{18} = 387.420.489$ different models.
Notice that this number does not include permutations
that lead to the same form for the Yukawa matrices.
But, it does include permutations
which, although leading to different forms of the
Yukawa matrices,
have no impact on the physical observables.
These same procedure must be followed when
we count the number of distinct forms of the Yukawa matrices
based on the analysis of the previous sections.

The forms shown in section~\ref{subsec:one} correspond
to $6_L \times (3+3+3+3+6+6+6+6)_{nR} \times (3+3+3+3+6+6+6+6)_{pR} = 7776$.
The sub-indices $L$, $nR$, and $pR$ correspond to the permutations
of rows, down-type columns, and up-type columns (respectively),
that lead to the same physics.
But, as in the $3^{18}$ possibilities above,
the counting has been performed so that
no two structures look the same.
The numbers in $(3+3+3+3+6+6+6+6)_{nR}$
correspond to the number of possibilities in
Eqs.~(\ref{odd1})--(\ref{odd8}), respectively.

To be specific,
let us look at Eq.~(\ref{odd1}).
Exchanging the first and second column leaves the form invariant.
This is counted as one structure.
However,
exchanging the third and first columns leads to a new structure.
So does an exchange between the third and second column.
There are thus three possibilities.
This explains the first ``3'' in $(3+3+3+3+6+6+6+6)_{nR}$.
The rest of the counting procedure follows the same lines.

The forms shown in section~\ref{subsubsec:4.1} correspond
to $1_L \times (1+3+3+1)_{nR} \times (1+3+3+1)_{pR} = 64$.
The forms shown in section~\ref{subsubsec:4.3}
with $\theta \neq \pi$ correspond
to $3_L \times (1+3+3+6+3+3)_{nR} \times (1+3+3+6+3+3)_{pR} = 1083$.
The forms shown in section~\ref{subsubsec:4.3}
with $\theta = \pi$ correspond
to $3_L \times (3+3+3)_{nR} \times (3+3+3)_{pR} = 243$.
The forms shown in section~\ref{subsubsec:4.2} correspond
to $3_L \times (1+3+3+6+3+3)_{nR} \times (1+3+3+6+3+3)_{pR} = 1083$.
Finally,
forms shown in section~\ref{subsubsec:special_2pi/3} correspond
to $6_L \times (3+3+3+6+3+3+3)_{nR} \times (3+3+3+6+3+3+3)_{pR} = 3456$.
There are thus $13.705$ distinct surviving possibilities.

This may seem like a large number,
but notice that we have eliminated $387.406.784$
a priori conceivable Yukawa structures.
The simple requirements of quarks with nonzero mass and
a CKM matrix which is not block diagonal provides
a drastic reduction in the number of possibilities.
Said otherwise,
the huge majority of Yukawa matrices consistent with
abelian symmetries do not survive simple experimental
constraints.
We should also point out that any two structures
which differ only by permutations of the rows
(simultaneously in $\Gamma$ and $\Delta$),
and/or by permutations of the columns of $\Gamma$,
and/or by permutations of the columns of $\Delta$
give exactly the same physics.
Permutations aside,
we are left with the $8+4+9+6+7=34$ possibilities
for the down-type Yukawa matrices shown in
Eqs~(\ref{odd1})--(\ref{odd8}),
(\ref{4.1-1})--(\ref{4.1-4}),
(\ref{4.3-g})--(\ref{4.3-a}),
(\ref{4.2.2-notpi})--(\ref{4.2.8-notpi}),
and (\ref{2p3.1})--(\ref{2p3.7}),
with similar structures for the up-type Yukawa matrices.
Combining appropriately,
we get
$8 \times 8 + 4 \times 4 + 9 \times 9 + 6 \times 6 + 7 \times 7 = 246$ overall models.
Those that differ only by $\Phi_1 \leftrightarrow \Phi_2$
will lead to the same physics.
Of those,
a few can be further excluded because they do not yield
any CP violation.
The possibility of spontaneous CP violation
will be addressed in section~\ref{sec:CPV}.

\section{\label{sec:important}Two important results}

\subsection{\label{subsec:important1}Most discrete symmetries have the same impact}

We have considered a symmetry in the scalar sector $S = \mbox{diag}\{1 , e^{i \theta}\}$. 
Of course, if the lagrangian is invariant under $S$, it is invariant under any power of $S$.
In this way, if $\theta = 2\pi/n$, then the $Z_n$ group is generated. If $\theta \neq
2\pi/n$, then one generates a discrete, but infinite, group. For simplicity we will 
refer to the $Z_n$ groups in what follows. 

We now turn to an important result from our previous analysis.
We know that choosing $\theta=2\pi/3$ or $\theta=2\pi/5$ leads to the same
Higgs potential.
Indeed,
any $\theta \neq 0, \pi$ leads to the same Higgs potential
as the continuous $U(1)$ Peccei-Quinn symmetry \cite{FS1}.
From this point of view,
applying any $Z_n\, (n \geq 3)$,
or even $U(1)$ is the same.
With the results presented in the previous section,
we see that this is no longer the case when the fermions are added.
As shown here,
the symmetry $Z_3$ allows Yukawa structures not allowed
for other $Z_n$.
Remarkably,
all $Z_n$ with $n \geq 4$ have the same impact
on the full Lagrangian,
even when fermions are introduced.

\subsection{\label{subsec:important2}Most discrete symmetries imply an accidental continuous symmetry}

The notation
$\theta \neq 2 \pi/3, \pi, 4 \pi/3$
used in Eqs.~(\ref{odd1})--(\ref{odd8})
means that the form of
the matrices shown is the most general consistent with
values of $\theta$ which differ from
$2 \pi/3$, $\pi$, and $4 \pi/3$.
But one should notice that the form of the matrices shown
are left invariant even if $\theta = 2 \pi/3, \pi, 4 \pi/3$.
The point is that, in general, for those special values
of $\theta$ these matrix forms are \textit{not the most general}
consistent with the symmetries.
For example,
Eq.~\eqref{odd1} is not the most general matrix consistent
with $\theta_{11}=\theta_{12}=\theta$,
$\theta_{13}= 2 \theta$\ \ $(\textrm{mod } 2 \pi)$
when $\theta=\pi$.
That form is shown in Eq.~\eqref{4.3-b}.
But one can see that, indeed,
Eq.~\eqref{odd1} is a particular case of Eq.~\eqref{4.3-b}.
So, Eqs.~(\ref{odd1})--(\ref{odd8}) are invariant
under the symmetry for all $\theta$,
but they are only the most general forms
consistent with the symmetry for
those symmetries where $\theta \neq 2 \pi/3, \pi, 4 \pi/3$.
The dedicated reader can check this explicitly by comparing these
forms with the forms presented for the special cases
$\theta=\pi$ and $\theta=2\pi/3$.

This has a very important consequence.
A matrix form which is invariant under the symmetry
for some value of $\theta \neq 2 \pi/3, \pi, 4 \pi/3$
will be invariant under the symmetry for all
values of $\theta$,
meaning that the Yukawa sector will be invariant under $U(1)$.
Since this is also true for the Higgs potential,
we conclude that,
for the cases in section~\ref{subsec:one}:
i) Imposing $Z_2$ on the scalars does not imply a larger symmetry,
neither in the Higgs sector, nor in the Yukawa sector;
ii) Imposing $Z_3$ on the scalars implies a continuous symmetry
in the Higgs sector, but not in the Yukawa sector;
iii) Imposing $Z_n,\ n \geq 4$ on the scalars implies a continuous symmetry,
both in the Higgs sector and in the Yukawa sector.

The other cases can be analyzed in a similar fashion.
For the cases in section~\ref{subsubsec:4.1}:
i) Imposing $Z_2$ on the scalars implies a continuous symmetry
in Yukawa the sector, but not in the Higgs sector;
ii) Imposing $Z_n,\ n \geq 3$ on the scalars implies a continuous symmetry,
both in the Higgs sector and in the Yukawa sector.
For the cases in sections~\ref{subsubsec:4.3} and \ref{subsubsec:4.2}:
i) Imposing $Z_2$ on the scalars does not imply a larger symmetry,
neither in the Higgs sector, nor in the Yukawa sector;
ii) Imposing $Z_n,\ n \geq 3$ on the scalars implies a continuous symmetry,
both in the Higgs sector and in the Yukawa sector.

\section{\label{sec:CPV}Spontaneous CP violation}

\subsection{\label{subsec:strict}Strict two Higgs doublet model}

Let us now look at the possible vacua of a theory with only two Higgs doublets and three
fermion generations, and their implications for
CP violation at the lagrangian level.
We are interested in implementations of discrete 
abelian symmetries, like $Z_n$, for which the scalar potential of Eq.~\eqref{VH2} can
be written as 
\begin{eqnarray}
V
&=&
m_{11}^2 \Phi_1^\dagger \Phi_1 + m_{22}^2 \Phi_2^\dagger \Phi_2
-  m_{12}^2 (\Phi_1^\dagger \Phi_2 + \textrm{h.c.} ) \nonumber\\[6pt]
&+&
\tfrac{1}{2} \lambda_1 (\Phi_1^\dagger\Phi_1)^2
+ \tfrac{1}{2} \lambda_2 (\Phi_2^\dagger\Phi_2)^2
+ \lambda_3 (\Phi_1^\dagger\Phi_1) (\Phi_2^\dagger\Phi_2) 
+
\lambda_4 (\Phi_1^\dagger\Phi_2) (\Phi_2^\dagger\Phi_1)
+
\tfrac{1}{2} \lambda_5 \left[(\Phi_1^\dagger\Phi_2)^2 + \textrm{h.c.} \right] ,
\label{VH1}
\end{eqnarray}
where all the parameters are real. We have included the soft-breaking parameter
$m_{12}^2$, taken to be real so that CP is not explicitly broken. For a $Z_2$ symmetry 
- $\theta = \pi$ in Eq.~\eqref{S-simple2} - 
the $\lambda_5$ coupling is present in the potential. For $Z_n$, $n\geq 3$, or indeed any other
value for $\theta$ different from $0$ or $\pi$, the symmetry sets
$\lambda_5$ to zero and the potential is indistinguishable from the Peccei-Quinn 
one~\cite{PQ}. At the minimum, the scalar fields develop vevs which we take to be given by,
without loss of generality
\be
\langle \Phi_1 \rangle = v_1 = u_1 \;\;\; , \;\;\; \langle \Phi_2 \rangle = v_2 = u_2 + i u_3 \,,
\ee
with all $u_i$ real. A vacuum with $u_3 \neq 0$ {\em may} lead to spontaneous CP violation 
(SCPV) in the scalar sector - however, the presence of a phase in the vacuum is no guarantee of SCPV. To verify 
whether SCPV occurs in the scalar sector, we must calculate the basis invariant quantities of ref.~\cite{BS}, 
which was done for all possible THDM scalar potentials in~\cite{FMNS}. 
The minimization conditions are given by $\partial V/\partial u_i = 0$, from which we obtain
\ba
0 &=& \left[m_{11}^2 + \lambda_1 u_1^2 + (\lambda_3 + \lambda_4) (u_2^2 + u_3^2) + 
\lambda_5 (u_2^2 - u_3^2)\right] u_1 
\,-\,m_{12}^2 u_2 \label{eq:m1}\\[6pt]
0 &=& \left[m_{22}^2 + \lambda_2 (u_2^2 + u_3^2) + (\lambda_3 + \lambda_4 + \lambda_5) u_1^2\right] u_2
\,-\,m_{12}^2 u_1 \label{eq:m2}\\[6pt]
0 &=& \left[m_{22}^2 + \lambda_2 (u_2^2 + u_3^2) + (\lambda_3 + \lambda_4 - \lambda_5) u_1^2 \right] u_3\,.\label{eq:m3}
\ea
From these we see that solutions with $u_3 = 0$ are always possible. There are several interesting
cases:
\begin{itemize}
\item $\theta = \pi$, exact $Z_2$ symmetry ($m_{12}^2 = 0$, $\lambda_5 \neq 0$): from 
Eqs.~\eqref{eq:m3} and~\eqref{eq:m2}, 
any solution with $u_3 \neq 0$ automatically implies either $u_1 = 0$ or $u_2 = 0$. 
Both solutions lead to no SCPV
in the scalar sector (see~\cite{FMNS}).
\item $\theta = \pi$, softly broken $Z_2$ symmetry ($m_{12}^2$, $\lambda_5 \neq 0$): 
both solutions without SCPV in the scalar sector ($u_3 = 0$)
and with SCPV in the scalar sector ($u_3 \neq 0$) are possible, depending on the values 
of potential's parameters~\cite{BR}.
\item $\theta \neq \{0, \pi\}$, exact $U(1)$ symmetry ($m_{12}^2 = \lambda_5 = 0$): 
the equations above only determine the
sum $u_2^2 + u_3^2$, and as such the relative phase of the vevs is arbitrary. These vacua lead to no 
SCPV in the scalar sector~\cite{FMNS} and in fact generate an axion.
\item $\theta \neq \{0, \pi\}$,  softly broken $U(1)$ symmetry ($m_{12}^2 \neq 0$, $\lambda_5 = 0$): 
from Eqs.~\eqref{eq:m3} and~\eqref{eq:m2},
we see that any solution with $u_3 \neq 0$ leads to $u_1 = 0$ which, considering Eq.~\eqref{eq:m1}, 
also implies $u_2 = 0$. Thus, no SCPV vacuum can occur in this case. Vacua with $u_3 = 0$ possess no axion. 
\end{itemize}
The existence of an axion in one of the cases above is easy to understand: as was explained earlier, 
the imposition of a discrete symmetry with $\theta \neq \{0, \pi\}$ (for instance a $Z_n$ symmetry 
with $n\geq 3$) on the scalar potential leads to an accidental Peccei-Quinn continuous $U(1)$
symmetry. Any vacuum for which both fields acquire a vev will break that symmetry and lead to a zero mass for the pseudoscalar. This corresponds in fact to the appearance of an additional Goldstone boson (other than the three 
usual ones arising from the breaking of the gauge symmetry). Analytically, the pseudoscalar mass is given by
\be
m^2_A \,=\,\frac{v^2}{u_1 u_2}\, m_{12}^2\,+\,2 \lambda_5 v^2 \, ,
\ee
with $v^2 = u_1^2 + u_2^2$, for vacua with $u_3 = 0$~\footnote{In the case of the exact $U(1)$ symmetry 
an arbitrary phase between the vevs is possible, but it has no effect on the scalar masses whatsoever.}. 
From this we see that: the $Z_2$ potential will never lead to an axion, since $\lambda_5 \neq 0$; the exact
$U(1)$ symmetry forces this mass to be zero; and the softly broken $Z_n$ potential again has no axion, as the
pseudoscalar mass is directly proportional to the soft breaking parameter. 

The scalar vevs originate the fermion masses, but also have a contribution to CP breaking at the
lagrangian level, whether they are real or complex. In fact, the Jarlskog invariant, which measures
CP violation in the weak interactions, is given by \cite{Ja}
\be
J = \textrm{Tr} [H_u,H_d]^3 = 6 i (m_t^2 - m_c^2)(m_t^2 - m_u^2)(m_c^2 - m_u^2) 
 (m_b^2 - m_s^2)(m_b^2 - m_d^2)(m_s^2 - m_d^2)
\textrm{Im} \left( V_{us} V_{cb} V_{ub}^\ast V_{cs}^\ast \right),
\label{eq:J}
\ee
where the matrices $H_d$ and $H_u$ have been defined in Eqs.~\eqref{H_d}
and~\eqref{H_u}. In the SM, since no CP breaking can arise spontaneously,
it is explicitly broken with complex Yukawa couplings. In the THDM we can study models
where one has demanded that the full lagrangian
be CP invariant, such that the matrices $\Gamma_i$ and $\Delta_i$
will be real, and the only possibility
of producing a non-zero Jarlskog invariant will be the vevs having a relative
phase.
Since such a vacuum is impossible for the softly broken $U(1)$ scalar potential,  
we conclude that models with an abelian symmetry (other than $Z_2$) and 
with an explicit CP conservation are ruled out,
since for them $J$ would always be zero.
Nonetheless, there is a distinction
worth making : the special forms found for the matrices with $\theta = 2\pi/3$ ($Z_3$ symmetry), 
given in section~\ref{sec:2pi3}, would give a non-zero Jarlskog invariant {\em if a vacuum with
a complex phase could be produced}; all the other Yukawa matrices we have obtained for the cases
$\theta \neq \pi, 2\pi/3$ give $J = 0$ {\em even if a complex vacuum existed}. As such, the only 
models allowed are those, like
the SM, where CP is explicitly broken by the Yukawa couplings.

As for the $Z_2$ model, the {\em exact} 
symmetry is also ruled out when CP is explicitly preserved - no phase from the vevs can originate 
$J \neq 0$, even for the odd case $u_2 = 0$, allowed by Eqs.~\eqref{eq:m1}--~\eqref{eq:m3}: in that 
case there is a phase of $\pi/2$ in the vaccum, but it has no bearing on $J$, which gives zero.
In the softly broken $Z_2$ model a vacuum with a relative phase in the vevs may
be obtained and it leads to CP violation, both in the scalar and the Yukawa sectors~\cite{BR}. 
And as before, $Z_2$ models with explicit CP
breaking are in principle perfectly viable. We summarize this analysis in Table~\ref{tab1}.
%
\begin{table*}[ht!]
\caption{Possibilities of CP violation for THDM with abelian symmetries.
``Yes" means that the model's parameters can generate a non-zero value for the
Jarlskog invariant. The ``$U(1)$" models are those for which one has imposed
a discrete symmetry of the form of Eq.~\eqref{S-simple2}, with $\theta\neq 0, \pi$.}
\begin{tabular}{ccc}
\hline \hline
Model & Lagrangian with & CP-conserving  \\
 & explicit CP breaking & Lagrangian \\
\hline
 & & No - real vacuum \\
Exact $Z_2$ &  Yes &  or   \\
 & & vev phase gives $J = 0$ \\
 \hline
 & & \\
Soft-broken $Z_2$ & Yes & Yes  \\
 & & \\
 \hline
  & & \\
 Exact $U(1)$ & Yes & No - vacuum \\
  & & gives axion \\
 \hline
 &  & \\
Soft-broken $U(1)$ & Yes & No - vacuum with phase\\
  & & impossible  \\  
\hline \hline
\end{tabular}
\label{tab1}
\end{table*}
%

A few observations are in order:
\begin{itemize}
\item We have not considered in this analysis the so-called ``inert vacua",
where either $\langle \Phi_1\rangle = 0$ or $\langle \Phi_2\rangle = 0$, possible
in the case of exact symmetries ($Z_2$ or $U(1)$). These give an 
acceptable $J$ only in the case of explicit CP breaking. 
\item The $Z_3$ case is special. Let us again consider the case of
explicit CP conservation. Unlike the remaining symmetries with $\theta \neq
\pi, 2\pi/3$, a vacuum with complex vevs would give $J\neq 0$. Such a 
vacuum is impossible in the THDM, but one can conceive (like the authors
of~\cite{BGL} did) models with two doublets and additional gauge singlets,
capable of producing the desired form for the vevs~\cite{FLS}.
\end{itemize}

\subsection{\label{subsec:complex}Complex vacua and the Jarlskog invariant}

The vacua of a $Z_n$ potential may be easily altered by
introducing soft-breaking terms,
as discussed in the previous section,
or by the inclusion of extra singlet scalars.
Here we discuss those cases where the introduction of
singlet scalars implies a relative phase between $v_1$
and $v_2$,
and we ask whether this provokes the appearance of
a phase in the CKM matrix when all Yukawa couplings
are real~\footnote{Of
course, the inclusion of scalar gauge
singlets has no impact on the Yukawa matrices we have found in the
previous sections, since singlet scalars have no coupling to the fermions.
}.

To do this we calculated the Jarlskog invariant of Eq.~\eqref{eq:J},
assuming a relative phase between $v_1$ and $v_2$ for all the
$246$ models of Yukawa matrices (assumed real) which we have identified.
In almost all cases $J=0$.
The only exceptions occur for $\theta=\pi$ or $\theta=2 \pi/3$.
The results are presented in Table~\ref{tab2} and Table~\ref{tab3},
respectively.
These tables will be useful for the study of spontaneous CP violation
in models with two scalar doublets and various scalar singlets,
in the presence of abelian symmetries.
%
\begin{table*}[ht!]
\caption{We assume that $\theta= \pi$,
that all Yukawa entries are real,
that the vevs have a relative complex phase,
and we calculate $J$.
The down-type Yukawas were chosen according to the equations along
the first line,
and the up-type Yukawas were chosen according to the equations along
the first column.
We denote the entries where $J=0$,
all others allow for $J \neq 0$,
depending on the values of the parameters.}
\begin{tabular}{|c|c|c|c|c|}
\hline \hline
Equations  &   &   &   &  \\
for Yukawa & (\ref{4.3-g}) & (\ref{4.3-d}) & (\ref{4.3-b}) & (\ref{4.3-a})\\
matrices   &   &   &   &  \\
\hline
(\ref{4.3-g}) & 0 &   &   &   \\
\hline
(\ref{4.3-d}) &   &   &   &   \\
\hline
(\ref{4.3-b}) &   &   &   &   \\
\hline
(\ref{4.3-a}) &   &   &   & 0 \\
\hline \hline
\end{tabular}
\label{tab2}
\end{table*}
%
\begin{table*}[ht!]
\caption{We assume that $\theta= 2 \pi/3$,
that all Yukawa entries are real,
that the vevs have a relative complex phase,
and we calculate $J$.
The down-type Yukawas were chosen according to the equations along
the first line,
and the up-type Yukawas were chosen according to the equations along
the first column.
We denote the entries where $J=0$,
all others allow for $J \neq 0$,
depending on the values of the parameters.}
\begin{tabular}{|c|c|c|c|c|c|c|c|}
\hline \hline
Equations  &   &   &   &   &   &   &  \\
for Yukawa
& \eqref{2p3.1} & \eqref{2p3.2} & \eqref{2p3.3} &
\eqref{2p3.4} & \eqref{2p3.5} & \eqref{2p3.6} & \eqref{2p3.7} \\
matrices  &   &   &   &   &   &   &  \\ 
\hline
\eqref{2p3.1} & 0 &   & 0 &   &   &   &   \\
\hline
\eqref{2p3.2} &   & 0 &   &   & 0 &   &   \\
\hline
\eqref{2p3.3} & 0 &   & 0 &   &   &   &   \\
\hline
\eqref{2p3.4} &   &   &   &   &   &   &   \\
\hline
\eqref{2p3.5} &   & 0 &   &   & 0 &   &   \\
\hline
\eqref{2p3.6} &   &   &   &   &   & 0 & 0 \\
\hline
\eqref{2p3.7} &   &   &   &   &   & 0 & 0 \\
\hline \hline
\end{tabular}
\label{tab3}
\end{table*}

\section{\label{sec:FCNSI}Natural suppression of Flavour Changing Neutral Scalar Interaction}

Measurements in the mixing of neutral mesons
(such as $K-\bar K$, $B_d - \bar B_d$, etc.)
lead to tight constraints on flavour changing neutral scalar
interactions (FCNSI).
The discrete symmetry $Z_2$ was introduced in the scalar sector
by Glashow and Weinberg \cite{Glashow:1976nt} and,
independently, by Paschos \cite{Paschos:1976ay},
precisely to preclude such FCNSI.
But there are several other options to curtail FCNSI.
For example,
one may invoke large scalar masses,
or introduce approximate flavour symmetries \cite{Rasin}.
Perhaps more interestingly,
one may relate the FCNSI with the CKM matrix.
In a very nice article,
Branco, Grimus, and Lavoura (BGL) used discrete abelian
symmetries in order to construct one such THDM \cite{BGL},
following earlier work by Lavoura \cite{Lav92}.
The BGL model corresponds to the use
of our Eq.~\eqref{4.3-e} for the up-type Yukawa matrices and
of our Eq.~\eqref{4.3-g} for the down-type Yukawa matrices.

One may now ask the question:
is there any other implementation of abelian symmetries which
leads to a relation between FCNSI and the CKM matrix?
Although we have all possible implementations of
abelian symmetries, the question is difficult to
answer analytically because it involves diagonalizing the mass matrices.
Indeed,
the quark mass basis is obtained with the basis transformation
\begin{eqnarray}
d_L &=& V_{dL}\ n_L,
\nonumber\\
d_R &=& V_{dR}\ n_R,
\nonumber\\
u_L &=& V_{uL}\ p_L,
\nonumber\\
u_R &=& V_{uR}\ p_R,
\label{mass-basis}
\end{eqnarray}
where we have used $q_L = (n_L, p_L)^\top$.
The unitary matrices $V_{dL}, V_{dR}, V_{uL}$, and $V_{uR}$
are chosen such that
\begin{eqnarray}
\textrm{diag} \{ m_d, m_s, m_b \} = D_d
& = &
V_{dL}\, \left[ v_1 \Gamma_1 + v_2 \Gamma_2 \right]\, V_{dR}^\dagger,
\nonumber\\
\textrm{diag} \{ m_u, m_c, m_t \} = D_u
& = &
V_{uL}\, \left[ v_1^\ast \Delta_1 + v_2^\ast \Delta_2 \right]\, V_{uR}^\dagger.
\label{DdDu}
\end{eqnarray}
The CKM matrix is $V=V_{uL} V_{dL}^\dagger$.
The matrices controlling the FCNSI are
\begin{eqnarray}
N_d
&=&
V_{dL}\, \left[ v_2^\ast \Gamma_1 - v_1^\ast \Gamma_2 \right]\, V_{dR}^\dagger,
\nonumber\\
N_u
&=&
V_{uL}\, \left[ v_2 \Delta_1 - v_1 \Delta_2 \right]\, V_{uR}^\dagger.
\label{NdNu}
\end{eqnarray}

Botella, Branco, and Rebelo \cite{Botella:2009pq}
have proposed a method to identify BGL-type
implementations while sidestepping the
diagonalization procedure.
They start from the relation \cite{Lav92}
\begin{equation}
N_d = \frac{v_2^\ast}{v_1} D_d - \frac{v^2}{v_1} V_{dL}\, \Gamma_2\, V_{dR}^\dagger,
\end{equation}
obtained by combining Eqs.~\eqref{DdDu} and \eqref{NdNu},
and using $v^2 = |v_1|^2 + |v_2|^2$.
Based on this they propose the following \textit{sufficient}
conditions for BGL implementation:
i) $ v_1^\ast \Delta_1 + v_2^\ast \Delta_2$ is block diagonal;
and ii) there exists a matrix $P$ such that
iia) $P \Gamma_2 = k \Gamma_2$ (for some number $k$),
and iib) $P \Gamma_1 = 0$.
As they stress,
the condition can be applied with an up-type/down-type quark
interchange.

We start by noticing that Eqs.~\eqref{DdDu} and \eqref{NdNu}
can also be combined into
\begin{equation}
N_d = - \frac{v_1^\ast}{v_2} D_d + \frac{v^2}{v_2} V_{dL}\, \Gamma_1\, V_{dR}^\dagger,
\end{equation}
implying that an equally good \textit{sufficient}
conditions for BGL implementation is:
i) $ v_1^\ast \Delta_1 + v_2^\ast \Delta_2$ is block diagonal;
and ii) there exists a matrix $P$ such that
iia) $P \Gamma_1 = k \Gamma_1$ (for some $k$),
and iib) $P \Gamma_2 = 0$.
Again,
the condition can be applied with an up-type/down-type quark
interchange.
The new condition is just a $\Phi_1 \leftrightarrow \Phi_2$ transformation
of the previous condition,
useful to us when looking for all possible BGL-type implementations.

Since we have tabled all possible matrices,
we are able to see that only Eq.~\eqref{4.3-e} can lead to
a block diagonal $ v_1^\ast \Delta_1 + v_2^\ast \Delta_2$ for
the up-type quarks.
We must now check all compatible down-type Yukawa matrices,
namely,
Eqs.~\eqref{4.3-g},
\eqref{4.3-h},
\eqref{4.3-i},
\eqref{4.3-c},
\eqref{4.3-f},
and see whether they satisfy condition ii)~\footnote{The possibility that
both the up-type and down-type Yukawa matrices
are given by Eq.~\eqref{4.3-e} is excluded,
since it would lead to a block-diagonal CKM matrix.
}.
We have checked that only for Eq.~\eqref{4.3-g} can one find
a matrix $P$ consistent with the constraints ii).

This gives a unique character to the work of Branco,
Grimus, and Lavoura \cite{BGL}.
They have developed the \textit{only} possible implementation
of a relation between FCNSI and the CKM matrix which uses
abelian symmetries and is consistent with the \textit{sufficient}
conditions above.
There are only two caveats.
First,
we have only checked the sufficient conditions developed
by Ref.~\cite{Botella:2009pq} and extended here.
A priori,
one can entertain the possible existence of cases
which \textit{do not} satisfy the sufficient conditions
presented,
but where the FCNSI are indeed related to the CKM matrix.
In the cases where we could perform the analysis analytically,
we have found no such case.
Second,
in some cases condition ii) is violated  because it leads
to constraints on the non-zero matrix elements of the Yukawa matrices.
It could be that some non-abelian group might lead to further zeros
on the Yukawa matrices, thus evading the problem.
Although possible,
such a case would be difficult to construct because
more zeros in the Yukawa matrices will,
more often than not,
lead to massless quarks or to a block-diagonal CKM matrix.

In light of our analysis,
that a BGL \cite{BGL} case was found by inspection
in the THDM is truly remarkable.

\section{\label{sec:conclusions}Conclusions}

We have studied the restrictions on the Yukawa matrices
imposed by discrete abelian symmetries acting
on the scalar and fermion sectors of the THDM.
Using known experimental constraints,
we have reduced the number of possible cases
from $3^{18}$ to 246.
Ignoring row and column permutations,
we are left with 34 types of down-type Yukawa matrices
(and the same for up-type quarks),
which we table explicitly.

We have found that imposing a symmetry $Z_n\ (n \geq 4)$
on the scalars always leads to an accidental $U(1)$ symmetry;
that applying a $Z_3$ symmetry on the scalars leads to
an accidental $U(1)$ symmetry in the
scalar sector but not necessarily
in the fermion sector;
and that applying a $Z_2$ symmetry on the scalars does not lead
to an accidental $U(1)$ symmetry in either sector.

We show that only $Z_2$ with soft breaking in the scalar sector
enables spontaneous CP violation.
We also show that the proposal of Branco, Grimus and Lavoura \cite{BGL}
is unique,
in our context,
and conjecture that this uniqueness might hold even when
non-abelian symmetries are considered in the THDM.

Finally, we stress that our results have a very wide applicability in model building
because all discrete non-abelian groups have
a $Z_n$ subgroup,
for some value of $n$.
For a given non-abelian group,
pick one of its $Z_n$ subgroups and diagonalize its generator.
Applying that generator as a symmetry of the lagrangian,
one falls into one of the 34 Yukawa matrices we have shown explicitly.
The action of further generators
(which, of course, need not be diagonalizable in the same basis)
will, in general,
lead to further constraints on the Yukawa matrices.
Given the low number of
entries in many of our Yukawa matrices,
and the likelyhood of further constraints setting them to zero,
the action of further generators will often lead to matrices
inconsistent with experimental constraints.


\begin{acknowledgments}

The work of P.M.F. is supported in part by the Portuguese
\textit{Funda\c{c}\~{a}o para a Ci\^{e}ncia e a Tecnologia} (FCT)
under contract PTDC/FIS/70156/2006.
The work of J.P.S. is funded by FCT through the projects
CERN/FP/109305/2009 and  U777-Plurianual,
and by the EU RTN project Marie Curie: MRTN-CT-2006-035505.
We are grateful to L.\ Lavoura for many discussions and suggestions,
and for reading this manuscript.

\end{acknowledgments}

\end{document}